\documentclass[aps,twocolumn,superscriptaddress,floatfix,nofootinbib]{revtex4-1}
\usepackage{graphicx,amsmath,amssymb,verbatim,color}
\usepackage{booktabs}
\usepackage{comment}
\usepackage{soul}
\usepackage[dvipsnames]{xcolor}
\usepackage{bm}
\usepackage[utf8]{inputenc}
\usepackage[colorlinks=true,citecolor=blue,linkcolor=blue,urlcolor=blue, backref=false,pdfborder={0 0 0}]{hyperref}
\usepackage{float}
\usepackage{multirow}
\newcommand{\be}{\begin{equation}\begin{gathered}}
\newcommand{\ee}{\end{gathered}\end{equation}} 
\newcommand{\barr}{\begin{eqnarray}}
\newcommand{\earr}{\end{eqnarray}} 
\usepackage{romannum}
\usepackage[normalem]{ulem}

\newcommand{\bs}{\boldsymbol}

\begin{document}

\pagenumbering{arabic}

\title{
What it takes to solve the Hubble tension \\
through modifications of cosmological recombination}
\author{Nanoom Lee}
\email{nanoom.lee@nyu.edu}
\affiliation{Center for Cosmology and Particle Physics, Department of Physics, New York University, New York, New York 10003, USA} 
\author{Yacine Ali-Ha\"imoud}
\email{yah2@nyu.edu}
\affiliation{Center for Cosmology and Particle Physics, Department of Physics, New York University, New York, New York 10003, USA} 
\author{Nils Sch\"oneberg}
\email{nils.science@gmail.com}
\affiliation{Institut de Ci\`encies del Cosmos, Universitat de Barcelona, Mart\'{\i} i Franqu\`es 1, Barcelona 08028, Spain}

\author{Vivian Poulin}
\email{vivian.poulin@umontpellier.fr}
\affiliation{Laboratoire Univers \& Particules de Montpellier, CNRS \& Université de
Montpellier (UMR-5299), 34095 Montpellier, France}

\date{\today} 
\begin{abstract}
We construct data-driven solutions to the Hubble tension which are perturbative modifications to the fiducial $\Lambda$CDM cosmology, using the Fisher bias formalism. Taking as proof of principle the case of a time-varying electron mass and fine structure constant, and focusing first on Planck CMB data, we demonstrate that a modified recombination \emph{can} solve the Hubble tension and lower $S_8$ to match weak lensing measurements. Once baryonic acoustic oscillation and uncalibrated supernovae data are included, however, it is not possible to fully solve the tension with perturbative modifications to recombination.
\end{abstract}
\maketitle

\textit{Introduction}.---The standard $\Lambda$ cold dark matter ($\Lambda$CDM) model has been providing an astonishing fit to a wide variety of cosmological data. Yet, the precise value of a very basic parameter of the model, the present-time expansion rate of the Universe (or Hubble constant) $H_0$, remains the subject of intense debate. On the one hand, this parameter can be inferred indirectly from early-Universe probes, under the assumption of standard physics and $\Lambda$CDM cosmology. The most precise early-Universe measurement is that inferred from the Planck satellite's Cosmic Microwave Background (CMB) anisotropy data, $H_0=67.36\pm0.54$ km/s/Mpc \cite{Planck2018}. On the other hand, $H_0$ can be obtained from local (or late-Universe) measurements. The most precise local measurement is that provided by the SH0ES Collaboration, which directly measures the current expansion rate from supernovae, using Cepheid variables as calibrators, arriving at $H_0 = 73.04 \pm 1.04$ km/s/Mpc \cite{Riess:2021jrx, Riess:2022mme}. Depending on the specific datasets considered, the discrepancy between early-Universe and local measurements has reached $4{\rm -}6\sigma$ \cite{Verde:2019ivm,Riess:2019qba}, enough to make this ``Hubble tension" one of the most pressing issues in recent cosmology.

Although one cannot exclude multiple unknown systematic errors as a reason for the discrepancy \cite{Rigault:2014kaa,Follin:2017ljs,Jones:2018vbn,NearbySupernovaFactory:2018qkd,Brout:2020msh,Efstathiou:2020wxn,Dainotti:2021pqg,Mortsell:2021nzg,Mortsell:2021tcx,Dainotti:2022bzg,Wojtak:2022bct}, it may also hint at new physics, or extensions of the $\Lambda$CDM model. To resolve this Hubble tension, an enormous number of models have been proposed. Late-time solutions, which include late dark energy, emergent dark energy, interacting dark energy, and decaying CDM, have been shown to be less effective \cite{DiValentino:2016hlg,Xia:2016vnp,DiValentino:2017zyq,Poulin:2018zxs,Knox:2019rjx,Arendse:2019hev,Benevento:2020fev,Camarena:2021jlr,Efstathiou:2021ocp,McCarthy:2022gok}. This is because postrecombination solutions do not change the sound horizon at baryon decoupling, $r_d$, and can therefore not fit baryonic acoustic oscillation (BAO) data and uncalibrated type \Romannum{1}a supernovae (SN\Romannum{1}a) data while increasing the Hubble constant (so-called ``sound horizon problem"). This implies that a modification in early-time cosmology is needed to solve the Hubble tension \cite{Bernal:2016gxb,Evslin:2017qdn,Aylor:2018drw,Arendse:2019hev,DiValentino:2021izs} (see also Ref.~\cite{Bernal:2021yli} for a newly proposed quantity in the context of $H_0$ tension: the age of the Universe, and see Ref.~\cite{Ivanov_20} for a study of the degeneracy of $H_0$ with the CMB monopole temperature $T_0$). Early-time solutions focus on the reduction of the sound horizon at recombination, through either an increase in energy density, e.g., via early dark energy (EDE) \cite{Karwal:2016vyq,Poulin:2018cxd,Lin:2019qug,Smith:2019ihp,Murgia:2020ryi,Kamionkowski:2022pkx} or additional dark radiation \cite{Blinov:2020hmc,Aloni:2021eaq,Schoneberg:2022grr}, or a modification of the recombination history itself by, for example, introducing primordial magnetic fields (PMFs) \cite{Jedamzik:2011cu, Jedamzik:2018itu, Jedamzik:2020krr} (see Refs.~\cite{Thiele:2021okz, Rashkovetskyi:2021rwg} for whether small-scale baryon clumping due to PMF can resolve the Hubble tension together with Ref.~\cite{Lee:2021bmn} for a general formalism to estimate the effect of small-scale baryon perturbations on CMB anisotropies) or varying fundamental constants \cite{Hart:2017ndk,Hart:2019dxi,Sekiguchi:2020teg,Hart:2021kad} (see also Refs.~\cite{Chiang:2018xpn,Liu:2019awo} for nonstandard recombination). Reducing the size of the sound horizon at recombination naturally leads to larger $H_0$ value, in order to keep the angular size of the sound horizon measured at ${\cal O}(0.1\%)$ precision with Planck CMB data unaffected \cite{Knox:2019rjx}. However, none of the proposed solutions have been robustly detected in the variety of cosmological data, and, further, the reduction of the Hubble tension is partly due to an increased uncertainty. See Refs.~\cite{DiValentino:2021izs,Schoneberg:2021qvd} for a recent summary and comparison of proposed models.

In this Letter, we move beyond the model-by-model approach as an effort to resolve the Hubble tension, and for the first time, make use of the Fisher-bias formalism to find minimal \emph{data-driven} extensions to the $\Lambda$CDM model producing desired shifts in cosmological parameters (in this case, an increase in $H_0$), while not worsening the fit to a given dataset. We cast this question as a well-defined simple mathematical problem. With this formalism, as examples, we extract the shape of a time-varying electron mass $m_e(z)$ or fine structure constant $\alpha(z)$ modifications that would result in a better agreement of a given early-Universe dataset with SH0ES.
Let us stress that our primary goal is not to find a compelling physical solution to the tension, i.e., that can easily be realized via a simple theoretical model [for modeling of $\alpha(z)$ see, e.g., Refs.~\cite{Sigurdson:2003pd,Brzeminski:2020uhm}]. Instead, we focus on establishing whether such solutions \emph{exist}, which is already a nontrivial question. Indeed, it could very well be that the relevant observables are only sensitive to a few integrals of the ionization history, and that arbitrary modifications of the latter could only project on a limited subspace of observable variations.

In short, we show that while one can find \emph{small} time-varying perturbations to $m_e$ or $\alpha$ that would entirely solve the Hubble tension between Planck and SH0ES, once BAO and uncalibrated SN\Romannum{1}a data are included, one can only lower the Hubble tension down to $\sim 2.4\sigma$ with perturbative modifications to recombination, not being able to entirely resolve the tension.

\textit{Setting up the problem}.---We denote general observables by $\bs{X}$, a vector which may contain multiple observables, such as CMB angular power spectra, BAO measurements, or any other. Specifically, we denote the observed data by $\bs{X}^{\rm obs}$ and the corresponding theoretical prediction by $\bs{X}(\vec{\Omega})$, where $\vec{\Omega} \equiv \{\omega_c, \omega_b, H_0, \tau, \ln(10^{10}A_s), n_s\}$ is a set of cosmological parameters. One can obtain the best-fit parameters by maximizing the likelihood of the data $\mathcal{L}[\bs{X}(\vec{\Omega}); \bs{X}^{\rm obs}]$, or equivalently minimizing $\chi^2 \equiv - 2 \ln \mathcal{L}$. Importantly, the best-fit parameters $\vec{\Omega}_{}$ and best-fit chi-squared $\chi^2_{\rm BF} \equiv \chi^2[\bs{X}(\vec{\Omega}_{\rm BF}); \bs{X}^{\rm obs}]$ both depend on the underlying theoretical model $\bs{X}(\vec{\Omega})$. In particular, if we consider a model $\bs{X}'(\vec{\Omega}) = \bs{X}(\vec{\Omega}) + \Delta \bs{X}(\vec{\Omega})$ that differs from the standard model $\bs{X}(\vec{\Omega})$ by a small amount $\Delta \bs{X}(\vec{\Omega})$, the resulting best-fit parameters and chi-squared are shifted. 

More specifically, we will consider changes in the theoretical model resulting from perturbations to a smooth function $f(z)$ on which it depends. The resulting changes to the best-fit parameters $\Delta \vec{\Omega}_{\rm BF}[\Delta f(z)]$ and chi-squared $\Delta \chi^2_{\rm BF}[\Delta f(z)]$ are both \emph{functionals} of $\Delta f(z)$. Our general goal, then, is to find the smallest possible perturbations $\Delta f(z)$ allowing to shift the best-fit parameters to a target value $\vec{\Omega}_{\rm target}$, while not worsening the quality of fit\footnote{Strictly speaking, estimating the quality of the fit also involve the number of degrees of freedom (dof). Here we limit ourselves to requiring no change in $\chi^2$, given that the main goal of this work is to check the existence of solutions. Also, note that the number of dof for an arbitrary function, which can be estimated, for example, by principal component analysis, will be anyway dominated by the large number of CMB data points, independently of the model.}. In other words, we want to solve the following constrained optimization problem, for different datasets and different functions $f(z)$:
\be
\textrm{minimize}(|| \Delta f||^2)  \ \ \textrm{with} \ \begin{cases} \vec{\Omega}_{\rm BF}[\Delta f(z)] = \vec{\Omega}_{\rm target}, \\[5pt]
\Delta \chi^2_{\rm BF}[\Delta f(z)] \leq 0,
\end{cases} \label{eq:optimization}
\ee
where $||\cdots||$ is the $L^2$ norm, 
$
|| \Delta f||^2 \equiv \int dz \left[\Delta f(z)\right]^2.
$
In principle, this optimization problem can be solved exactly if combined with Markov Chain Monte Carlo (MCMC) analysis (or a minimization process) to estimate $\vec{\Omega}_{\rm BF}[\Delta f(z)]$ and $\Delta \chi^2_{\rm BF}[\Delta f(z)]$ for each given $\Delta f(z)$. However, this exact method would be heavily computationally expensive. Hence, to keep the optimization problem tractable, we will first derive simple approximations for $\vec{\Omega}_{\rm BF}$ and $\chi^2_{\rm BF}$\,, relying on the Fisher approximation (e.g.~Ref.~\cite{Tegmark:1996bz}). We approximate the data as Gaussian distributed, with inverse-covariance matrix $\bm{M}=\bm{\Sigma}^{-1}$. In general, this matrix depends on $\bs{X}(\vec{\Omega})$ itself; we will denote $\bs{M}(\vec{\Omega}) \equiv \bs{M}[\bs{X}(\vec{\Omega})]$ for short. We define the chi-squared of a given cosmology $\vec{\Omega}$ as
\be
\label{eq:chi-square}
\chi^2(\vec{\Omega}) \equiv [\bm{X}(\vec{\Omega}) - \bm{X}^{\rm obs}] \cdot \bm{M}(\vec{\Omega}) \cdot [\bm{X}(\vec{\Omega}) - \bm{X}^{\rm obs}].
\ee
By Taylor expanding the chi-squared to second order around a fiducial cosmology $\vec{\Omega}_{\rm fid}$, which we assume to be reasonably close to the best-fit, and minimizing it, we find the (approximate) best-fit cosmology $\vec{\Omega}_{\rm BF}$,
\barr
\Omega^i_{\rm BF} &=& \Omega^i_{\rm fid} - \frac{1}{2}(F^{-1})_{ij} \frac{\partial \chi^2}{\partial \Omega^j}\Big|_{\rm fid},
\label{eq:bestOmega_X}
\earr
where $F$ is the Fisher matrix defined and approximated as
\be
F_{ij} \equiv \frac{1}{2}\frac{\partial^2\chi^2}{\partial \Omega^i \partial \Omega^j}\Big|_{\rm fid}\approx \left(\frac{\partial \bs{X}}{\partial \Omega^i} \cdot \bs{M} \cdot \frac{\partial \bs{X}}{\partial \Omega^j}\right)\Big{|}_{\rm fid}.
\label{eq:fishermat}
\ee
Provided that the fiducial model is sufficiently close to the observations, we can approximate $\partial \chi^2/\partial \Omega^i$ to include only the leading contribution, which then implies
\barr
\Omega^i_{\rm BF} &\approx& \Omega^i_{\rm fid} \nonumber\\
&&- (F^{-1})_{ij} \frac{\partial \bm{X}}{\partial \Omega^j}\Big|_{\rm fid} \cdot \bm{M}(\vec{\Omega}_{\rm fid})\cdot [\bm{X}(\vec{\Omega}_{\rm fid}) - \bm{X}^\text{obs}].~~~~\label{eq:Omega_bf_final}
\earr
Inserting this solution into the Taylor-expanded chi-squared we find the approximate best-fit chi-squared
\barr
\chi^2(\vec{\Omega}_{\rm BF}) \approx [\bm{X}(\vec{\Omega}_{\rm fid}) - \bm{X}^{\rm obs}] \cdot\widetilde{\bm{M}} \cdot [\bm{X}(\vec{\Omega}_{\rm fid}) - \bm{X}^{\rm obs}],~
\label{eq:bestchi2_X}
\earr
where $\widetilde{\bm{M}}$ is defined as
\be
\widetilde{M}_{\alpha\beta} \equiv M_{\alpha\beta} - M_{\alpha \gamma} \frac{\partial X^\gamma}{\partial \Omega^i}(F^{-1})_{ij}\frac{\partial X^\sigma}{\partial \Omega^j}M_{\sigma\beta},
\ee
where $\bs{M}$ and $\partial \bs{X}/\partial \Omega^i$ are evaluated at the fiducial cosmology, and repeated indices are to be summed over. A simple property of the matrix $\widetilde{\bm{M}}$ is that it admits $\partial \bs{X}/\partial \Omega^i$ as null eigenvectors. We can thus think of $\widetilde{\bm{M}}$ as the inverse-covariance matrix of the data after marginalization over shifts in standard cosmological parameters.

\textit{Introducing new physics}.---Our main results so far, Eqs.~\eqref{eq:Omega_bf_final} and \eqref{eq:bestchi2_X}, apply to an arbitrary theoretical model, provided that it gives a reasonable fit to the data for the chosen fiducial cosmological parameters $\vec{\Omega}_{\rm fid}$. The best-fit parameters and chi-squared of a new theoretical model $\bs{X}'(\vec{\Omega}) = \bs{X}(\vec{\Omega}) + \Delta \bs{X}(\vec{\Omega})$ differ from those of the standard model $\bs{X}(\vec{\Omega})$ by small amounts $\Delta \Omega_{\rm BF}^i$ and $\Delta \chi^2_{\rm BF}$, respectively. Assuming $\Delta \bs{X}(\vec{\Omega}_{\rm fid})$ and $\bs{X}(\vec{\Omega}_{\rm fid}) - \bs{X}^{\rm obs}$ are approximately of the same order of magnitude, and by writing a change in the theoretical model due to changes in a smooth function $f(z)$ as
\be
\Delta \bm{X} = \int dz\; \frac{\delta \bm{X}}{\delta f(z)} \Delta f(z)\label{eq:DX},
\ee
we obtain the resulting changes in the best-fit parameters and chi-squared
\barr
\Delta \Omega_{\rm BF}^i &=& \int dz\; \frac{\delta \Omega_{\rm BF}^i}{\delta  f(z)} \Delta f(z),\label{eq:DObf_X}\\
\Delta \chi_{\rm BF}^2 &=& \int dz\; \frac{\delta \chi_{\rm BF}^2}{\delta f(z)}\Delta f(z) \nonumber\\
&+& \frac12 \iint dz\;dz'\; \frac{\delta^2 \chi_{\rm BF}^2}{\delta f(z) \delta f(z')} \Delta f(z) \Delta f(z'),~~\label{eq:Dchi2bf_X}
\earr
where
\barr
\frac{\delta \Omega_{\rm BF}^i}{\delta f(z)} &=& -(F^{-1})_{ij} \frac{\partial \bm{X}}{\partial \Omega^j} \cdot \bm{M}\cdot\frac{\delta \bm{X}}{\delta f(z)},\label{eq:dObf}\\
\frac{\delta \chi_{\rm BF}^2}{\partial f(z)} &=& 2[\bm{X}(\vec{\Omega}_{\rm fid}) - \bm{X}^{\rm obs}] \cdot\widetilde{\bm{M}} \cdot \frac{\delta \bm{X}}{\delta f(z)},\label{eq:dchi2bf_lin}\\
\frac{\delta^2 \chi_{\rm BF}^2}{\delta f(z) \delta f(z')} &=& 2 \frac{\delta \bm{X}}{\delta f(z)}\cdot\widetilde{\bm{M}}\cdot\frac{\delta \bm{X}}{\delta f(z')},\label{eq:dchi2bf_quad}
\earr
where $F_{ij}, \bs{M}, \widetilde{\bs{M}}$, $\partial \bs{X}/\partial \Omega_i$ and $\delta \bs{X}/\delta f(z)$ are all to be evaluated at the fiducial cosmology and in the standard model. With the simplified expressions of Eqs.~\eqref{eq:DObf_X}-\eqref{eq:dchi2bf_quad}, the optimization problem of Eq.~\eqref{eq:optimization} becomes tractable. The equations above are known as the Fisher-bias formalism \cite{Knox:1998fp,Kim:2003mq,Taylor:2006aw,Shapiro:2008yk,DeBernardis:2008tk}, used in Refs.~\cite{Huterer:2000mj,Samsing:2009kb} to constrain arbitrary functions. While the formalism is well known, the \emph{application} we make of it is completely novel.

While our formalism is general and could be applied to any function $f(z)$ on which observables depend, in this Letter we will consider modifications to the cosmological ionization history. Specifically, we will consider time-dependent relative variations of the electron mass [$f(z) = \ln m_e(z)$] in the main text, generalizing the constant change to the electron mass which has been shown to be a promising solution \cite{Sekiguchi:2020teg,Hart:2019dxi,Schoneberg:2021qvd}. We also consider time-dependent variations of the fine structure constant [$f(z) = \ln \alpha(z)$], in Appendix.~\ref{appendix:me}.\footnote{The variations in the net recombination rate is another interesting possible extension we considered. However, it happens to inherit a stronger non-linearity of $C_\ell$'s, hence we do not include it in this Letter (see also Appendix.~\ref{appendix:error-approximations}).}

The functional derivatives $\delta \bs{X} / \delta f(z)$ are obtained numerically by adding narrow (Dirac-delta-like) changes to the smooth function $f(z)$, at different redshifts. This is done by modifying the recombination code \textsc{hyrec\nobreakdash-2} \cite{Ali-Haimoud:2010tlj, Ali-Haimoud:2010hou, Lee:2020obi} implemented in \textsc{class} \cite{class} (see Appendix.~\ref{Appendix:numerical-techniques} for details). This part of the calculation is similar to what has been done in principal component analyses (PCAs) of recombination perturbations \cite{Farhang:2011pt,Hart:2019gvj}. Despite this technical similarity, the mathematical problem we solve is very different from the one considered in PCAs, which search the eigenmodes of the (discretized) matrix $\delta^2 \chi^2/\delta f(z) \delta f(z')$ with the largest eigenvalues. In words, PCAs look for perturbations to recombination to which the data is \emph{most} sensitive, while in contrast, our goal is to find the \emph{smallest} perturbations producing a desired shift in best-fit cosmological parameters while \emph{not} increasing the best-fit $\chi^2$. See Appendix.~\ref{appendix:PCAs} for the differences in two analyses.

We will now apply this general formalism to Planck CMB anisotropy data and then to the combined Planck + BAO, Planck + BAO + PantheonPlus \cite{Brout:2022vxf} dataset, with the goal of finding data-driven solutions to the Hubble tension. Note that by BAO we denote BOSS DR12 anisotropic BAO measurements \cite{BOSS:2016wmc}.

\textit{Result I: Application to Planck CMB data}.---Here, the vector $\bs{X}$ consists of the binned lensed temperature and polarization power spectra, $\bm{X}\equiv \{ D_\ell^{\rm TT},D_\ell^{\rm TE},D_\ell^{\rm EE}\}$. For $\ell \geq 30$, we use the Planck-lite foreground-marginalized binned spectra and covariance matrix. For $\ell <30$, we adopt the compressed log-normal likelihood of Prince and Dunkley~\cite{Prince:2021fdv}, which has been shown to give virtually the same constraints as the exact low-$\ell$ Planck likelihood (and therefore we use $\bs{X} \equiv \{ \ln D_\ell^{\rm TT}, \ln D_\ell^{\rm TE},  \ln D_\ell^{\rm EE}\}$ for $\ell < 30$). We set our fiducial cosmology to the Planck best-fit $\Lambda$CDM parameters \cite{Planck2018}.

\begin{figure}[ht]
\includegraphics[width = \columnwidth,trim= 00 20 00 10]{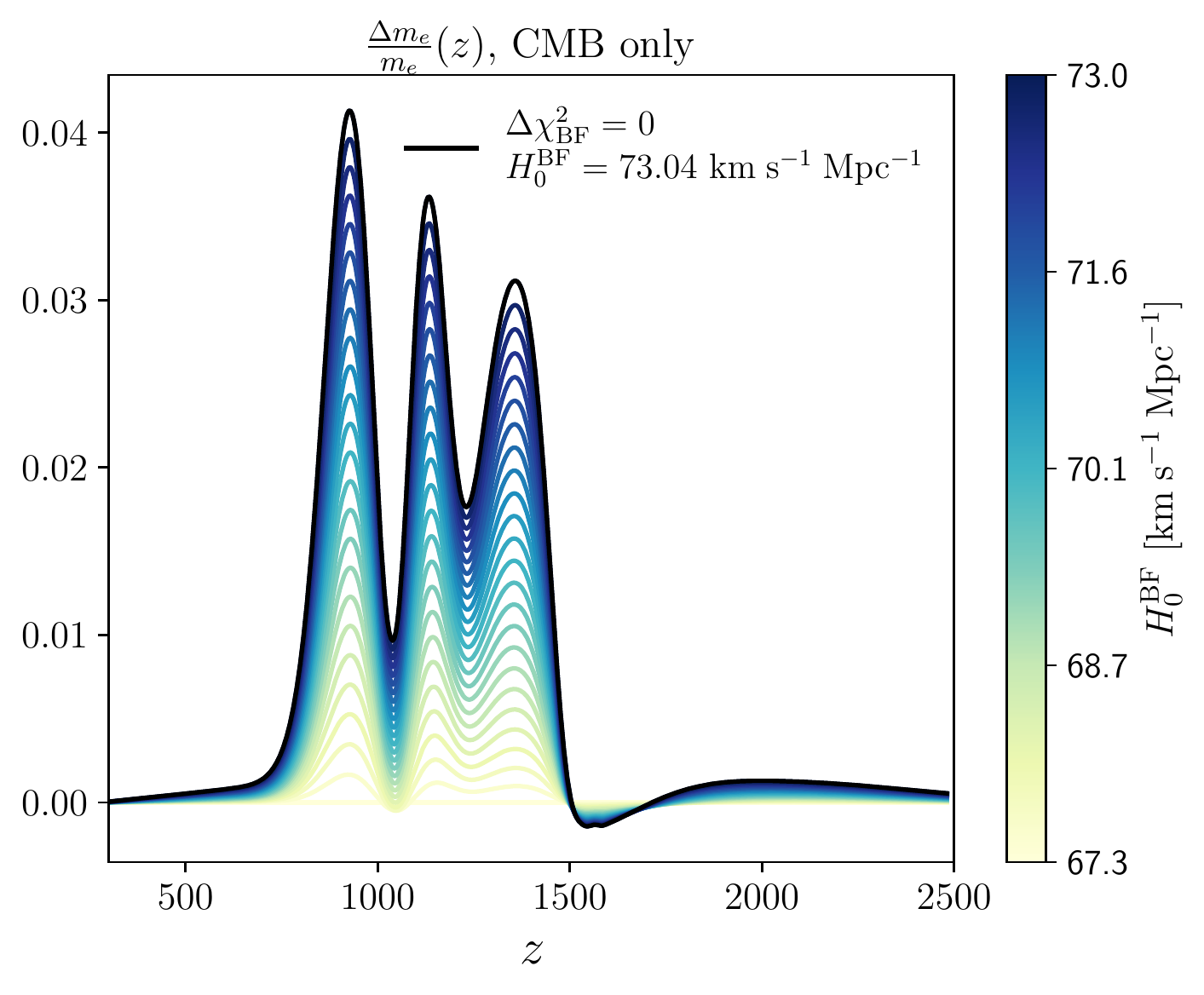}
    \caption{Solutions for $\frac{\Delta m_e}{m_e}(z)$ given target values of the CMB-only best-fit Hubble constant $H_0$, using Planck data \cite{Planck2018} alone. All solutions are constructed to keep the Planck best-fit chi-squared unaffected. The solution with a best-fit consistent with SH0ES \cite{Riess:2021jrx} $H_0^{\rm BF} = H_0^{\rm SH0ES}\equiv 73.04\;\text{km\;s}^{-1}\text{Mpc}^{-1}$(black curve), is denoted as $\Lambda$CDM + $m_e(z)$ model in the text.}
\label{fig:me-cmb}
\end{figure}

Using our formalism, we find variations of the time-varying electron mass $m_e(z)$ that cause the value inferred from Planck CMB anisotropies to be equal to a given target Hubble constant $H_0^\mathrm{BF}$ while not deteriorating the best fit chi-squared $\Delta \chi^2_{\rm BF}\leq0$. These deviations of $m_e(z)$ from its standard value are shown in Fig.~\ref{fig:me-cmb} with a range of $H_0^\mathrm{BF}$ values whose upper bound is the recent SH0ES best-fit \cite{Riess:2021jrx}. It is striking that our solution exhibits three large oscillations offset from zero between $z\simeq 700{\rm -}1500$. This behavior is significantly different than what has been modeled in past literature, namely, either a constant shift in $m_e$ or a power-law dependence on redshift \cite{Hart:2017ndk,Hart:2019dxi,Schoneberg:2021qvd}, explaining why these studies did not find as good a solution as we do, and illustrating the power of our formalism which can further be used to guide model-building (e.g., \cite{Brzeminski:2020uhm}). In particular, such a solution would also avoid big bang nucleosynthesis (BBN) constraints \cite{Seto:2022xgx}. We keep it for future work to investigate possible physical mechanisms that may generate the required oscillations. We confirm that the obtained $m_e(z)$ does indeed result in the expected parameter shifts by performing a MCMC analysis using \textsc{montepython} v3.0 \cite{Audren:2012wb,Brinckmann:2018cvx} with the full Planck TT, TE, EE + low E + lensing likelihood (see Appendix.~\ref{appendix:validation} for this validation test). This also attests that our use of the high-$\ell$ Planck-lite Gaussian likelihood combined with the low-$\ell$ compressed likelihood of Ref.~\cite{Prince:2021fdv}, as well as our noninclusion of the lensing potential likelihood, is accurate enough to derive a solution $m_e(z)$. In Appendix.~\ref{appendix:error-approximations}, we further quantify the accuracy of the other two approximations we make to derive the $m_e(z)$ solution, namely, the Taylor expansion of $\chi^2$ around a fiducial cosmology, and the linearity of observables in $\Delta f(z)$.

\begin{figure}[ht]
\includegraphics[width = .95\columnwidth,trim= 20 20 20 10]{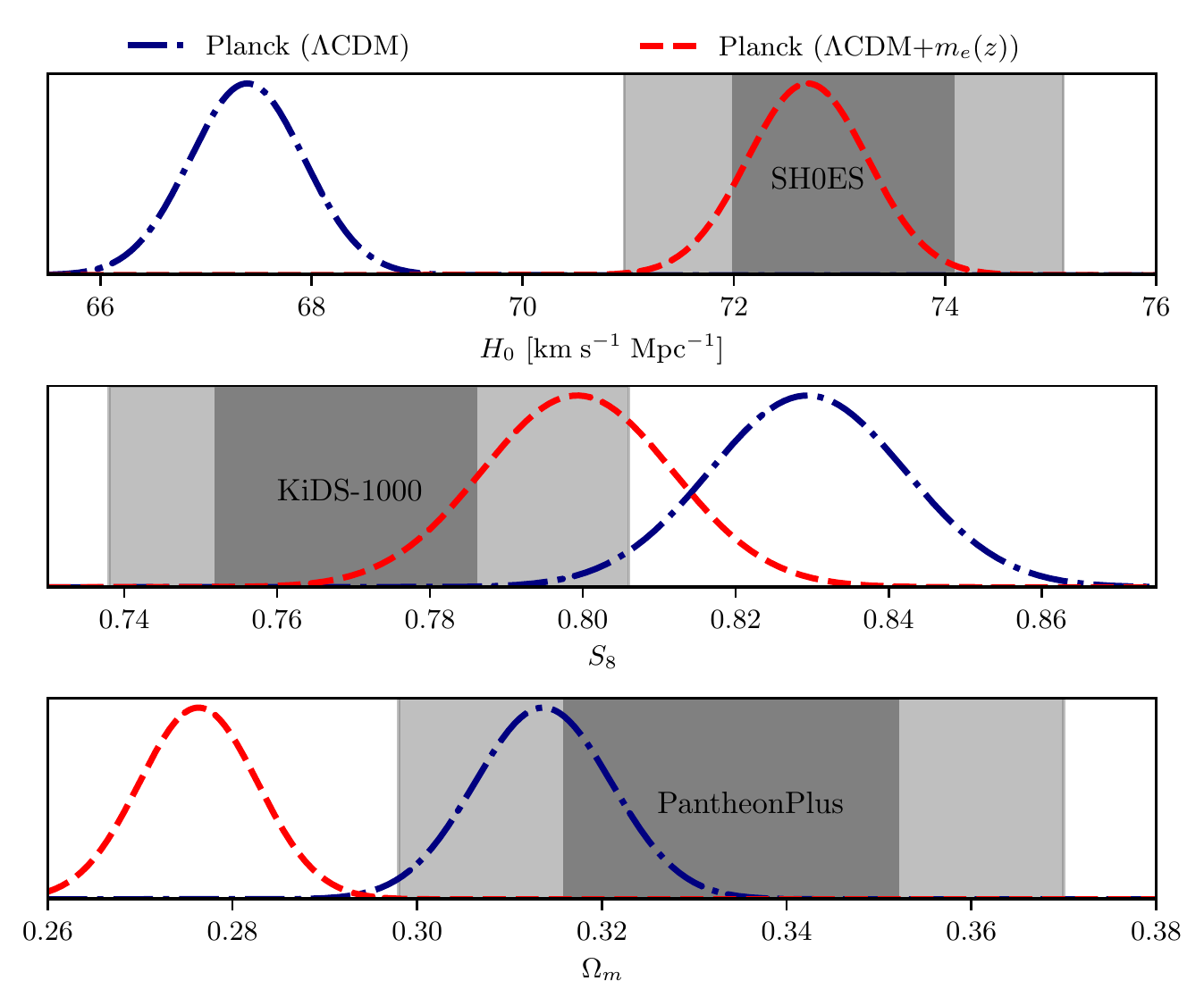}
\caption{Posteriors of $H_0$ (top), $S_8$ (middle), and $\Omega_m$ (bottom) inferred from Planck full likelihood, together with SH0ES, KiDS-1000, and PantheonPlus results, shown as grey bands.}
\label{fig:H0Om}
\end{figure}

Our main result is shown in the top panel of Fig.~\ref{fig:H0Om}, which displays the posterior distributions for $H_0$. In short, the $\Lambda$CDM model with a time-varying $m_e(z)$ given by the black curve in Fig.~\ref{fig:me-cmb} is a solution to the Hubble tension between Planck CMB and SH0ES data. It lowers the discrepancy to $0.29\sigma$ resulting in $H_0=72.70\pm0.57\;\text{km\;s}^{-1}\text{Mpc}^{-1}$, and $\Delta \chi^2=-1.15$ compared to the $\Lambda$CDM model\footnote{Throughout the Letter, the chi-squared from the chains is calculated at the mean cosmology assuming the best-fit cosmology is very close to the mean. This is due to difficulty of minimization process. The fact that all the posteriors of parameters are nearly Gaussian justifies this approximation.}. Interestingly, this solution tailored to solve the $H_0$ tension also happens to mostly solve another infamous tension in cosmology, the so-called $S_8$ tension ($S_8 = \sigma_8 \sqrt{\Omega_m/0.3}$), which is a discrepancy in measurements of the amplitude of matter clustering at the scale of 8 Mpc/$h$ between weak lensing probes and the value inferred from CMB anisotropies \cite{DES:2017myr,KiDS:2020suj,DiValentino:2020vvd}. Indeed, the $\Lambda$CDM + $m_e(z)$ model brings the Planck best-fit $S_8$ value to $1.1\sigma$ from the recent DES-Y3 constraint $S_8=0.776\pm0.017$ \cite{DES:2021wwk} and within $\sim 1.4\sigma$ from KiDS-450 ($S_8=0.745 \pm 0.039$) \cite{Hildebrandt:2016iqg} and KiDS-1000 ($S_8=0.766^{+0.020}_{-0.014}$) \cite{Heymans:2020gsg}, down from the $2$--$2.6\sigma$ tension in $\Lambda$CDM, as shown in the middle panel of Fig.~\ref{fig:H0Om}.

However, this extension to the $\Lambda$CDM model is less consistent with two other crucial cosmological data, BAO \cite{BOSS:2016wmc} and PantheonPlus \cite{Brout:2022vxf}. The bottom panel of Fig.~\ref{fig:H0Om} shows that this model is inconsistent with the PantheonPlus result for $\Omega_m$ with $\sim 3\sigma$ tension. This is fundamentally due to the well-known dependence of $\theta_s$ on $\Omega_m h^3$, which requires the best-fit $h$ and $\Omega_m$ to change in opposite ways, as we describe in further detail in Appendix \ref{app:change_me}. In addition, the agreement with BOSS DR12 BAO data is worsened resulting in an increase of the chi-squared of BOSS DR12 anisotropic measurements, $\Delta \chi^2_{\rm BAO} = +5.37$ (see also Ref.~\cite{Jedamzik:2020zmd} for a similar result\footnote{Note that, however, our results cannot be directly compared with those of Ref.~\cite{Jedamzik:2020zmd} due to the fixed relation between two sound horizon scales at baryon decoupling and recombination in Ref.~\cite{Jedamzik:2020zmd}, which is not satisfied in our case.}).

\textit{Result II: Application to Planck CMB + BAO or Planck CMB + BAO + PantheonPlus}.---We include either BAO or BAO + PantheonPlus data in our data vector $\bm{X}$, in order to see if we can obtain solutions to the Hubble tension that do not violate the agreement with these data that is present in the $\Lambda$CDM model. For BAO, we include BOSS DR12 anisotropic measurements at three effective redshifts $z_{\rm eff}=0.38, 0.51, 0.61$ \cite{BOSS:2016wmc},
$
\left\{\frac{D_M(z_{\rm eff})r_d^{\rm fid}}{r_d}, \frac{H(z_{\rm eff}) r_d}{r_d^{\rm fid}}\right\} \subset \bm{X},
$
and for PantheonPlus \cite{Brout:2022vxf} we include its constraint on the fractional energy density of the total matter \{$\Omega_m\} \subset \bm{X}$. We find that, in order to solve the Hubble tension either together with BAO or BAO + PantheonPlus, variations of $m_e(z)$ with larger amplitude are required together with larger shifts in best-fit cosmology. We note that these required more radical changes in recombination history and best-fit cosmology induce larger errors from the approximations taken in our formalism, preventing us from finding a self-consistent solution with a target $H_0^{\rm BF}=73.04\;\text{km\;s}^{-1}\text{Mpc}^{-1}$. Yet, we find that one can still \emph{partially} ease the Hubble tension with our current method while remaining within the regime of validity of our perturbative treatment. Explicitly, we find that one could lower the Hubble tension down to $\sim 1.3\sigma$ with BAO data included, while satisfying $\Delta \chi^2<1$ and $|\Omega_{\rm BF}^{\rm formalism} - \Omega_{\rm BF}^{\rm MCMC}|/\sigma<1$ for all six cosmological parameter $\Omega$'s. However, the reconstructed $\Omega_m \simeq 0.287$ is again less consistent with PantheonPlus compared to the standard $\Lambda$CDM model. Further, when PantheonPlus data are added along with CMB and BAO, we find that one could lower the tension down to at most $\sim2.4\sigma$ while remaining within the region of validity of the formalism. As such, even with arbitrary perturbative modifications to $\alpha(z)$ or $m_e(z)$ around the time of recombination, the Hubble tension can only be partially eased once BAO and supernovae data are accounted for.

\textit{Conclusions}.---We have built on the Fisher bias formalism to systematically search for data-driven solutions to any tension between any given datasets, by looking for the smallest possible change to an arbitrary function leading to a desired small shift in cosmological parameters\footnote{While we find minimal extensions not worsening the fit to a given data set, one could also seek for solutions to the tension with other strategies. For example, rather than aiming for a specific value of $H_0$, one can look for extensions minimizing the fit to all data sets including SH0ES/Pantheon/BAO (putting an additional constraint on the norm of the solution). We defer exploring these different strategies to future work.}. We applied our formalism to find a time-dependent function for the electron mass (and for the fine structure constant in Appendix.~\ref{appendix:me}) leading to a Hubble constant consistent with SH0ES while providing an equally good fit to Planck CMB data. We show that as a remarkable byproduct it happens to also solve the $S_8$ tension. However, this extended model is less consistent with BAO \cite{BOSS:2016wmc} and PantheonPlus \cite{Brout:2022vxf}.

Once BAO and PantheonPlus data are included in the formalism, we find that larger changes in recombination history are required to achieve the same target value of $H_0^{\rm BF}$, making the assumed linearity of observables, and the validity of Taylor expansion of $\chi^2$, break down. We note that these limitations can in principle be removed if one approaches the optimization problem with an exact method, which we defer to future work. In practice, we find that small perturbations to recombination through a time-varying electron mass can only reduce the tension down to $2.4\sigma$, and decreasing it further would likely require nonperturbative changes to recombination.

While we focus on perturbations to recombination in this Letter, our formalism can be applied more generally to any quantity impacting the prediction of a cosmological observable, e.g., the Hubble rate $H(z)$. We trust that the phenomenological framework we laid out, and the specific examples we provide here in terms of a modified recombination, will inspire a model-building effort from the cosmology and particle physics community with potential implications well beyond the mere study of cosmological tensions.

We thank Jens Chluba, Colin Hill, Marc Kamionkowski, Julien Lesgourgues, and Licia Verde for useful conversations. N.\;L. is supported by the Center for Cosmology and Particle Physics at New York University through the James Arthur Graduate Associate Fellowship. Y.\;A.\;H. is a CIFAR-Azrieli Global Scholar and acknowledges support from Canadian Institute for Advanced Research (CIFAR). N.\;S. acknowledges support from the Maria de Maetzu fellowship grant: CEX2019-000918-M, financiado por MCIN/AEI/10.13039/501100011033. V.\;P. is partly supported by the CNRS-IN2P3 grant Dark21. This project has received support from the European Union’s Horizon 2020 research and innovation program under the Marie Skodowska-Curie Grant Agreement No.~860881-HIDDeN and from COST Action CA21136 Addressing observational tensions in cosmology with systematics and fundamental physics (CosmoVerse) supported by COST (European Cooperation in Science and Technology). This project has received funding from the European Research Council (ERC) under the European Union’s HORIZON-ERC-2022 (Grant agreement No. 101076865).

\bibliography{mybib}

\pagebreak
\onecolumngrid

\begin{appendix}

\section{Data}
\subsection{CMB -- Planck}
As CMB data, for high-$\ell$'s ($\ell \geq 30$) we use Planck 2018 binned spectra (\texttt{cl\_cmb\_plik\_v22.dat}) and covariance matrix (\texttt{c\_matrix\_plik\_v22.dat}) with $\ell_{\rm max} = 2508$ for temperature and $\ell_{\rm max}=1996$ for polarizations, which is denoted as ``Planck-lite" \cite{Planck2018}. For low-$\ell$'s ($\ell < 30$), we adopt the compressed low-$\ell$ Planck likelihood constructed by Ref.~\cite{Prince:2021fdv}\footnote{\href{https://github.com/heatherprince/planck-low-py}{https://github.com/heatherprince/planck-low-py}}, where the likelihood for binned spectra $D_\ell \equiv \ell(\ell+1)C_\ell/2 \pi$ is given by
\be
\mathcal{L}(x) = p(x) = \frac{1}{(x-x_0) \sigma \sqrt{2\pi}} e^{-(\ln (x-x_0) - \mu)^2/(2\sigma^2)},
\ee
for $x=D_{\rm bin}$, with two and three bins for $TT$ and $EE$ spectrum, respectively. This is the best-fit log-normal probability distribution with values of $x_0$, $\mu$, and $\sigma$ determined in Ref.~\cite{Prince:2021fdv}.  We write the chi-squared from this likelihood as 
\be
\chi^2_{\text{low-}\ell} \equiv -2\ln \mathcal{L}(x) = \frac{ [\ln (x-x_0) - \mu + \sigma^2]^2}{\sigma^2} + \text{const}.
\ee
Ignoring the constant contribution, this chi-squared has the same form as that of Gaussian distributed data so that it can be included in our formalism which approximates the given data as Gaussian distributed.

\subsection{BAO -- BOSS DR12}
As BAO data, we use BOSS DR12 anisotropic measurements,
\be
\Bigg\{\frac{D_M(z_{\rm eff})r_d^{\rm fid}}{r_d},\frac{H(z_{\rm eff}) r_d}{r_d^{\rm fid}}\Bigg\},
\ee
at three effective redshifts $z_{\rm eff}=0.38~,0.51,~0.61$ \cite{BOSS:2016wmc}\footnote{\href{https://data.sdss.org/sas/dr12/boss/papers/clustering/ALAM_ET_AL_2016_consensus_and_individual_Gaussian_constraints.tar.gz}{https://data.sdss.org/sas/dr12/boss/papers/clustering/\\ALAM\_ET\_AL\_2016\_consensus\_and\_individual\_Gaussian\\\_constraints.tar.gz}}. The determination of the sound horizon at the drag epoch, $r_d$, in \textsc{class} \cite{class} is based on finding an exact point where the optical depth reaches unify, $\int_0^{\eta_d} d\eta\;a n_e \sigma_T /R=1$, where $\eta$ is the conformal time, $n_e$ is the number density of electrons, $\sigma_T$ is the Thomson cross section, and $R\equiv3\rho_b/4\rho_r$. While this determination is accurate enough for the standard recombination, it does not correctly reflect the effect of non-standard recombination scenarios on the sound horizon. Hence, instead, we determine the sound horizon scale as the location of the local maximum of the two-point correlation function $\xi(r)$, which is calculated from the linear matter power spectrum $P_m(k)$ using the Python package \textsc{cluster toolkit}\footnote{\href{https://github.com/tmcclintock/cluster_toolkit}{https://github.com/tmcclintock/cluster\_toolkit}} developed for Dark Energy Survey Year 1 stacked cluster weak lensing analysis \cite{DES:2018kma}. This determination gives $r_d^{\rm fid} \approx 149.22$ Mpc with BOSS DR12 fiducial cosmology. This fiducial sound horizon scale is different from that given in Ref.~\cite{BOSS:2016wmc}, which is $147.78$ Mpc. However, this is simply due to different conventions for defining $r_d$ and it has been known that the ratio of $r_d$ for different cosmologies are independent of the convention \cite{Aubourg:2014yra}.

\begin{figure}[ht]
\includegraphics[width = .32\columnwidth,trim= 00 10 00 20]{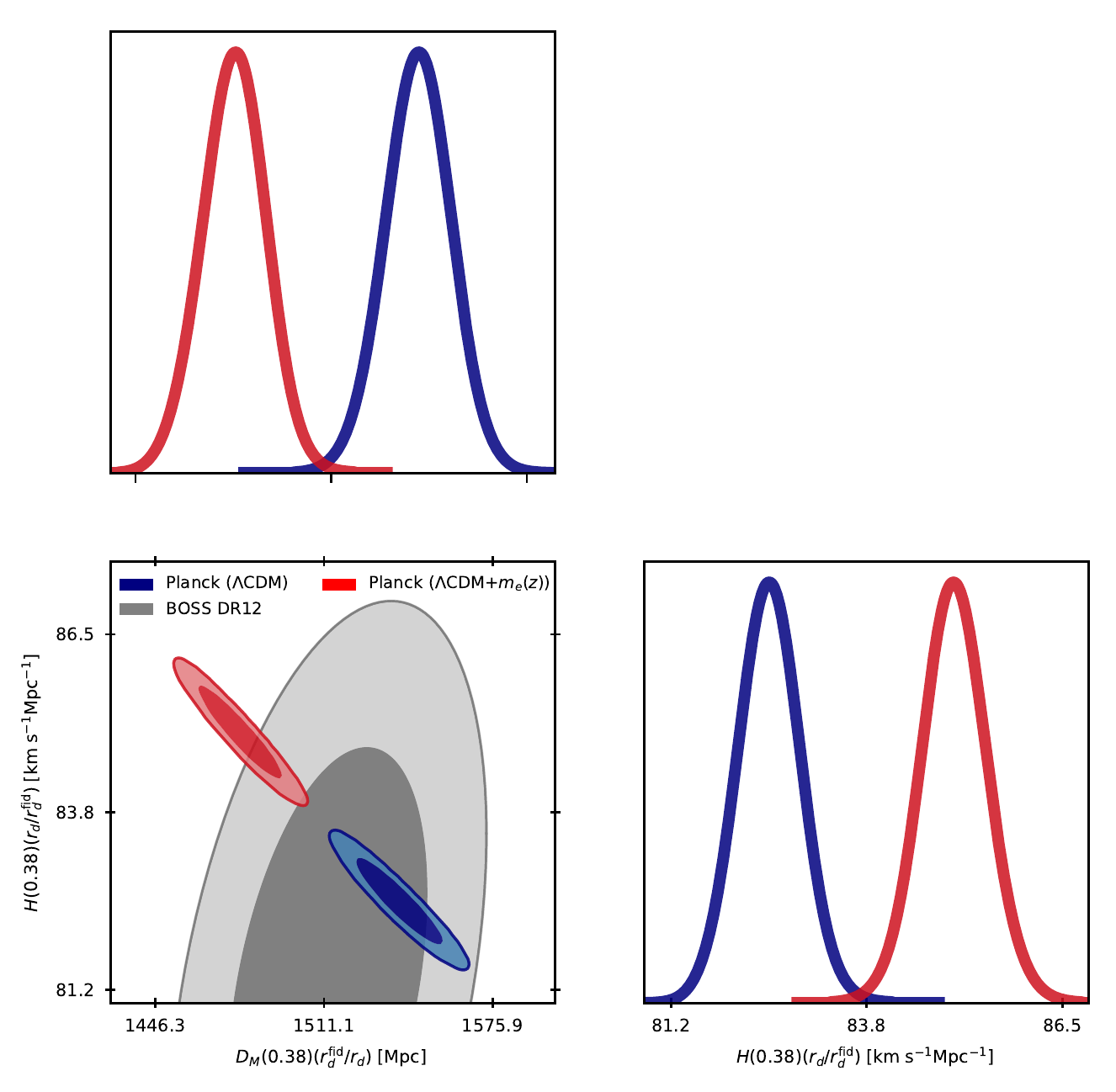}
\includegraphics[width = .32\columnwidth,trim= 00 10 00 20]{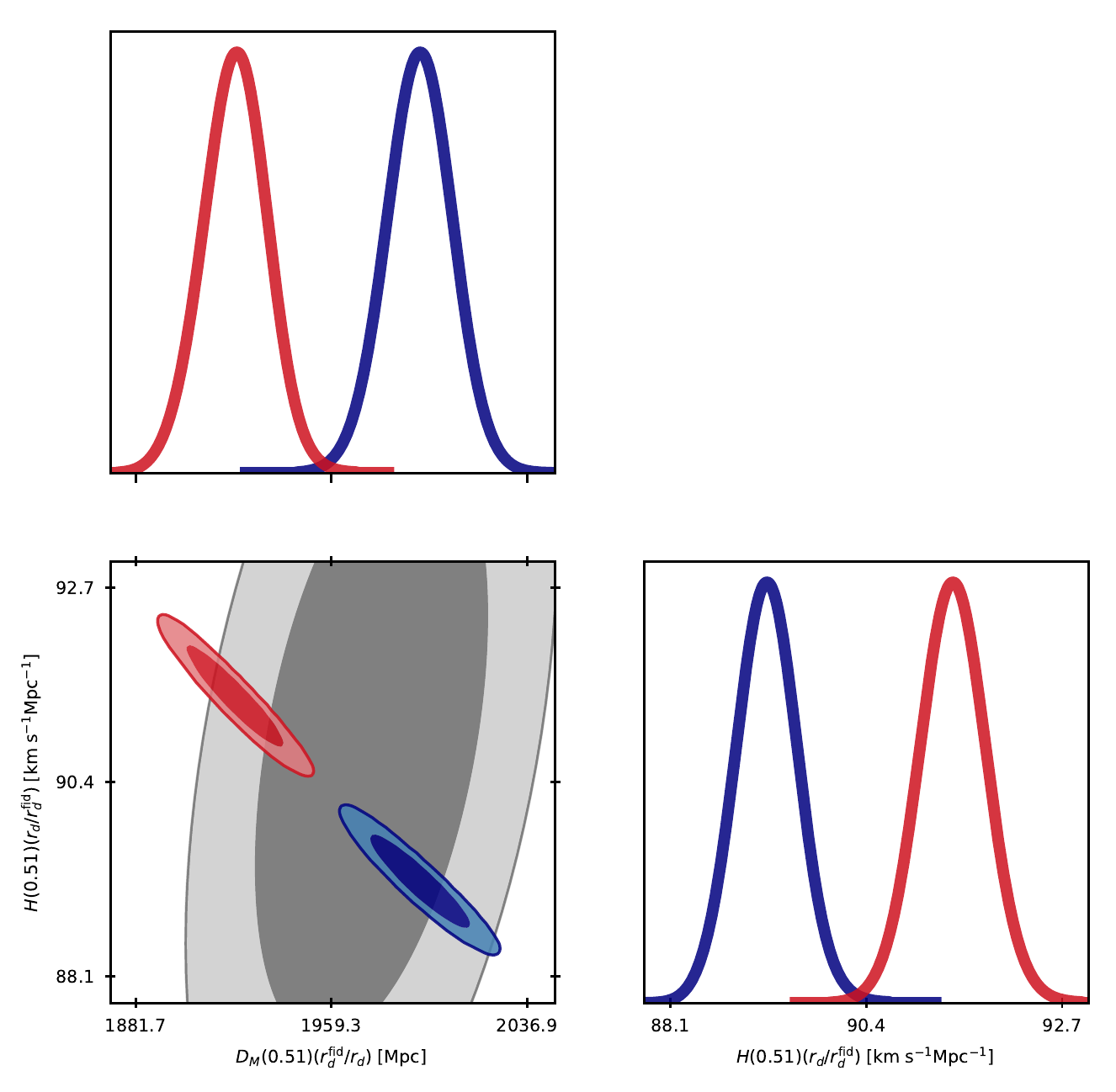}
\includegraphics[width = .32\columnwidth,trim= 00 10 00 20]{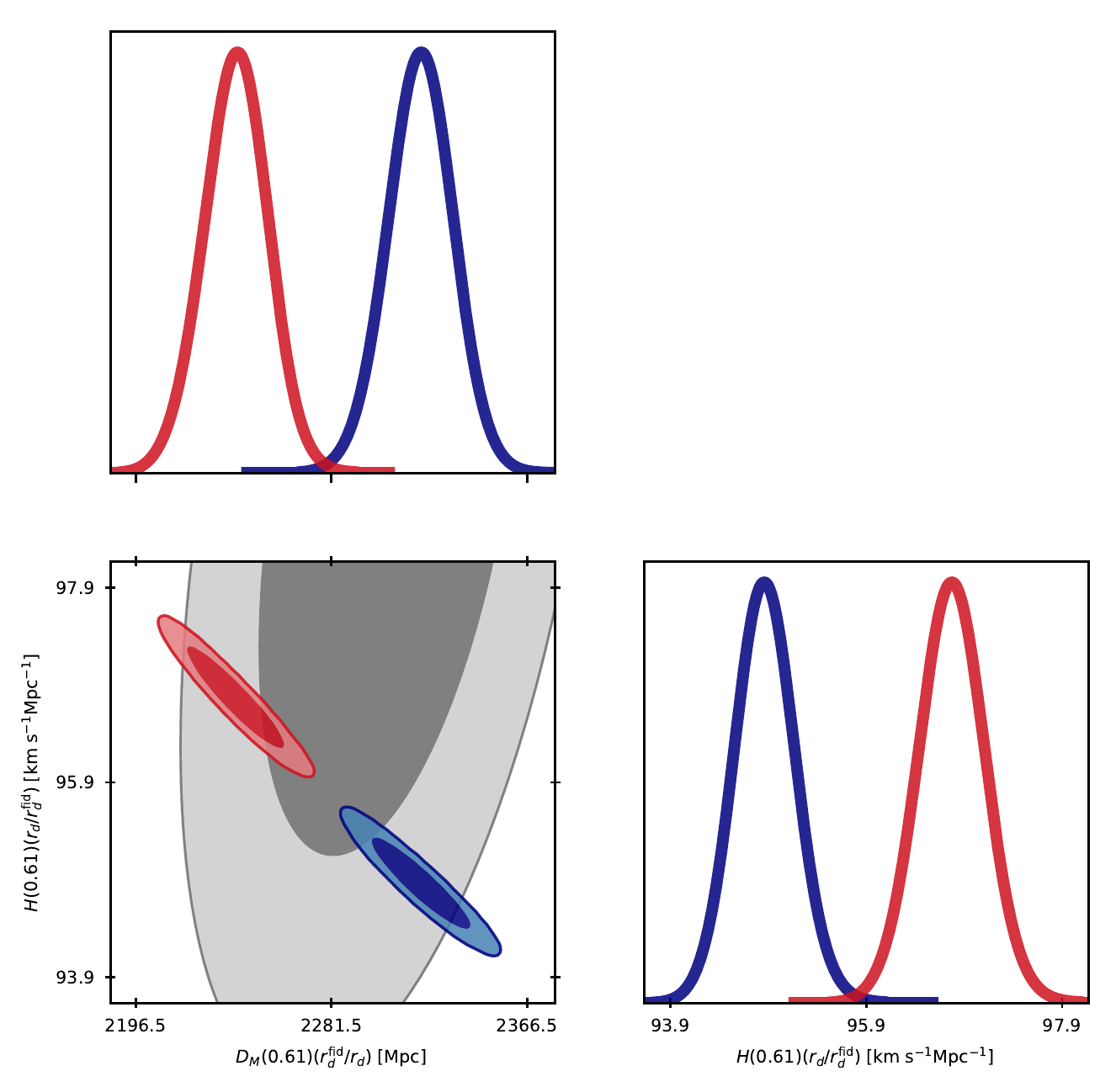}
\caption{68\% and 95\% confidence level constraints from BOSS DR12 \cite{BOSS:2016wmc} at three effective redshifts $z_{\rm eff}=0.38,0.51,0.61$ (grey). Other colored contours are the samples of each model with Planck full likelihood. The $\Lambda$CDM + $m_e(z)$ model resulting in larger $H_0$ value consistent with SH0ES \cite{Riess:2021jrx} is less consistent with BAO data compared to $\Lambda$CDM model ($\Delta \chi^2_{\rm BAO} = +5.37$), due to relatively larger $r_d$ compared to the value required to fit BAO data with such a high $H_0$ value.}
\label{fig:bao-}
\end{figure}

Figure~\ref{fig:bao-} shows the worsened agreement with BAO data of the $\Lambda$CDM + $m_e(z)$ model compared to the $\Lambda$CDM model.

\subsection{Uncalibrated SNIa -- PantheonPlus}

As an additional late-time observable data, we consider the constraint on the fractional energy density of total matter $\Omega_m = 0.334 \pm 0.018$ from PantheonPlus \cite{Brout:2022vxf}. We include this constraint in our formalism by rewriting $\Omega_m$ as a function of the cosmological parameters used in the formalism $\vec{\Omega} \equiv \{\omega_c, \omega_b, h, \tau, \ln(10^{10}A_s), n_s\}$, where $\omega_i \equiv \rho_i h^2/\rho_{\rm crit}$ are the dimensionless physical density parameters, and $h \equiv H_0$/(100 km/s/Mpc). Explicitly, we have 
\be
\Omega_m = (\omega_c + \omega_b + \omega_\nu) h^{-2},
\ee
where $\omega_\nu = 0.000644$ is fixed with one massive neutrino species of $m_\nu=0.06$ eV. We consider this constraint on $\Omega_m$ from PantheonPlus \cite{Brout:2022vxf} as an additional data by including it in the data vector $\bm{X}$.

\section{Numerical techniques}
\label{Appendix:numerical-techniques}
For the considered smooth function $f(z)$, as perturbations to this function, we define $N(=120)$ Dirac-delta-like functions in the range of redshifts $z_b(\equiv300)\leq z \leq z_e(\equiv2500)$ similarly to PCA in Ref.~\cite{Hart:2019gvj},
\be
\delta \ln f(z,z_i) \equiv \frac{\Delta f}{f}(z,z_i) = \frac{1}{\sqrt{2\pi\sigma^2}} \exp{\left[ - \frac{(z-z_i)^2}{2\sigma^2}\right]}, \quad \quad \sigma \equiv \frac{z_e-z_b}{6N\sqrt{2\ln2}}.
\ee
Using these functions, we calculate the functional derivatives at $N$ redshifts $z_i$'s and then interpolate to get $\delta \Omega^i_{\rm BF} / \delta \ln f(z)$ and $\delta \chi^2_{\rm BF}/\delta \ln f(z)$ as two-sided numerical derivatives by modifying \textsc{hyrec\nobreakdash-2} \cite{Ali-Haimoud:2010tlj, Ali-Haimoud:2010hou, Lee:2020obi} and \textsc{class} \cite{class}. For time-varying electron mass and fine structure constant, we use the dependencies of the energy levels of hydrogen and helium, atomic transition rates, photo-ionization/recombination rates summarized in Refs.~\cite{Kaplinghat:1998ry,Scoccola:2009xtv,Planck:2014ylh,Chluba:2015gta,Hart:2017ndk}, which had been already implemented in \textsc{hyrec\nobreakdash-2} \cite{Ali-Haimoud:2010tlj, Ali-Haimoud:2010hou, Lee:2020obi} as shown below.
\barr
\mathcal{A}_{2s},\;\mathcal{A}_{2p} ~&\propto&~ \alpha^2m_e^{-2}\label{eq:A2sA2p}\\
\mathcal{B}_{2s},\;\mathcal{B}_{2p},\;\mathcal{R}_{2p2s},\;\mathcal{R}_{2s2p} ~&\propto&~ \alpha^5m_e\\
\Lambda_{2s,1s} ~&\propto&~ \alpha^8m_e\\
\sigma_T ~&\propto&~ \alpha^2m_e^{-2}\label{eq:sigmaT}\\
T_\text{eff} ~&\propto&~ \alpha^{-2}m_e^{-1}\label{eq:Teff}
\earr
For the definitions and expressions of those quantities, see Refs.~\cite{Lee:2020obi,Ali-Haimoud:2010hou,Ali-Haimoud:2010tlj}.

\section{Changes in Planck's best-fit and chi-squared due to time-varying electron mass $m_e(z)$}
\label{app:change_me}

\begin{figure}[ht]
\centering
\includegraphics[width = .9\columnwidth,trim= 00 20 00 20]{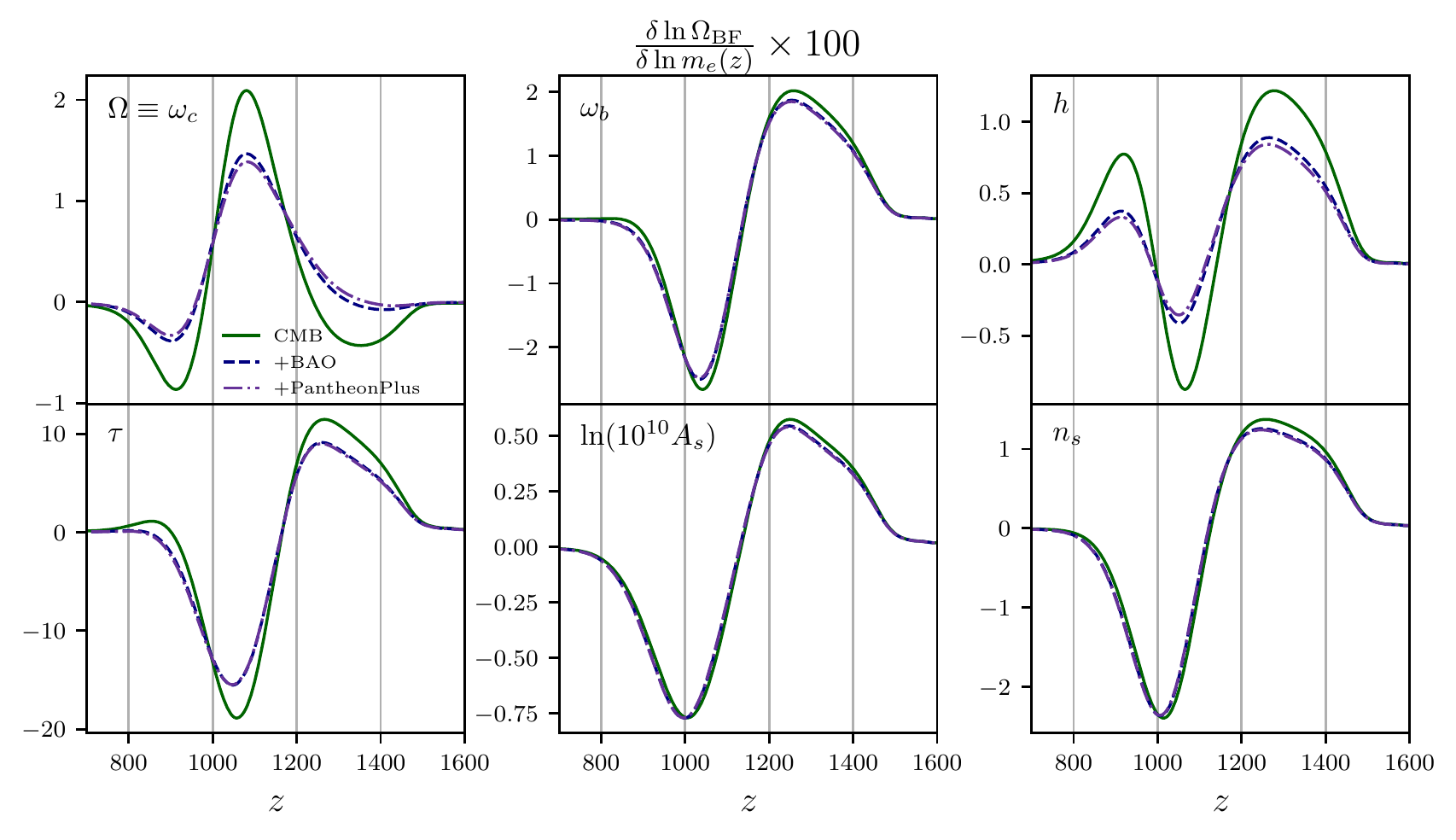}
\caption{Functional derivatives of best-fit parameters, Eq.~\eqref{eq:dObf} with three data sets (Planck CMB, Planck CMB + BOSS DR12 BAO, Planck CMB + BOSS DR12 BAO + PantheonPlus). Note that the range of redshifts we consider for perturbations in $m_e$ is $z=[300,2500]$, although here we plot with narrower range of redshifts for better presentation, since changes in recombination at $z \gtrsim 1500$ or $z \lesssim 800$ have little effect on CMB anisotropy power spectra. Note that $h \equiv H_0/(100$ km/s/Mpc).}
\label{fig:deriv-Omega}
\end{figure}

\begin{figure}[ht]
\centering
\includegraphics[width = .4\columnwidth,trim= 10 10 00 20]{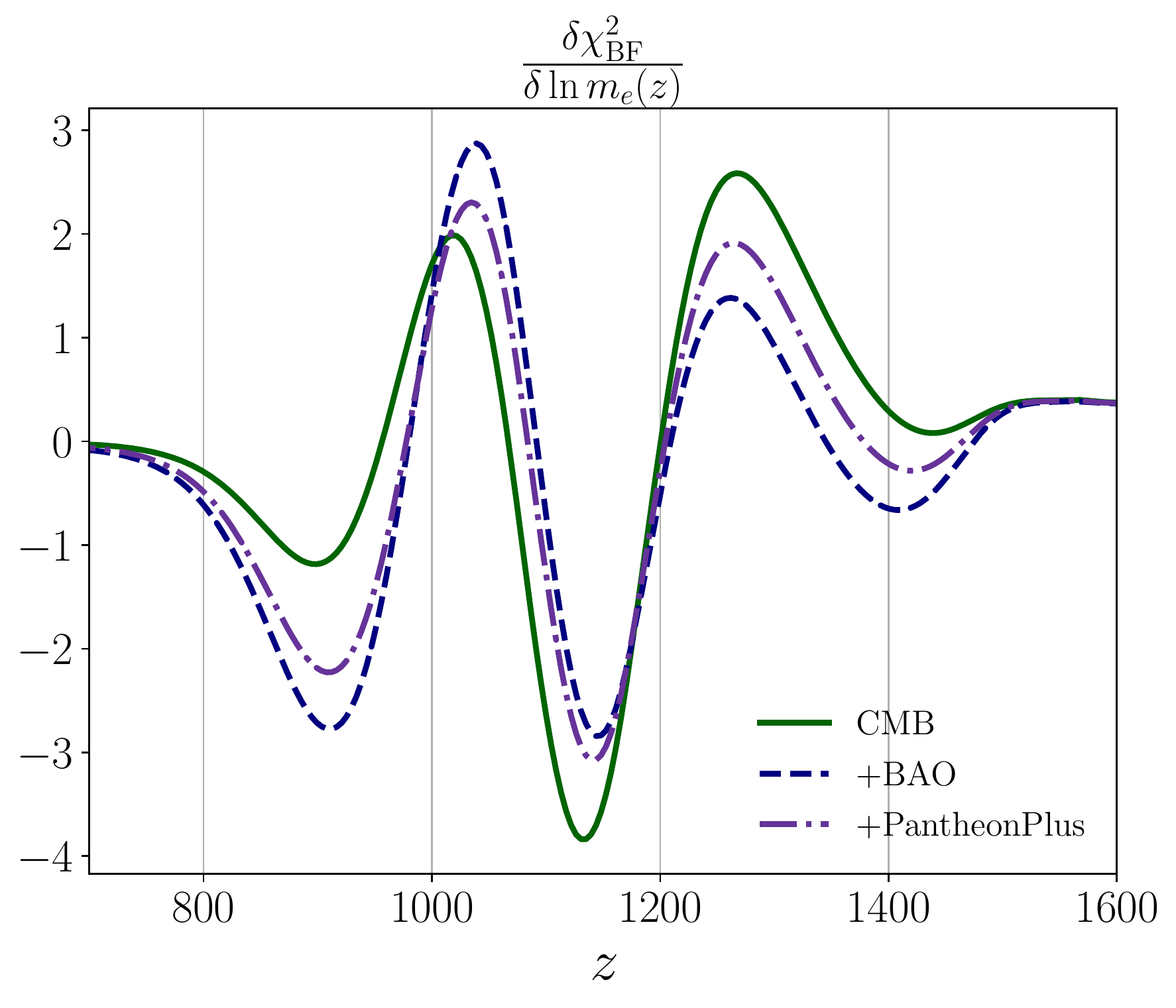}
\\\includegraphics[width = .32\columnwidth,trim= 10 10 00 00]{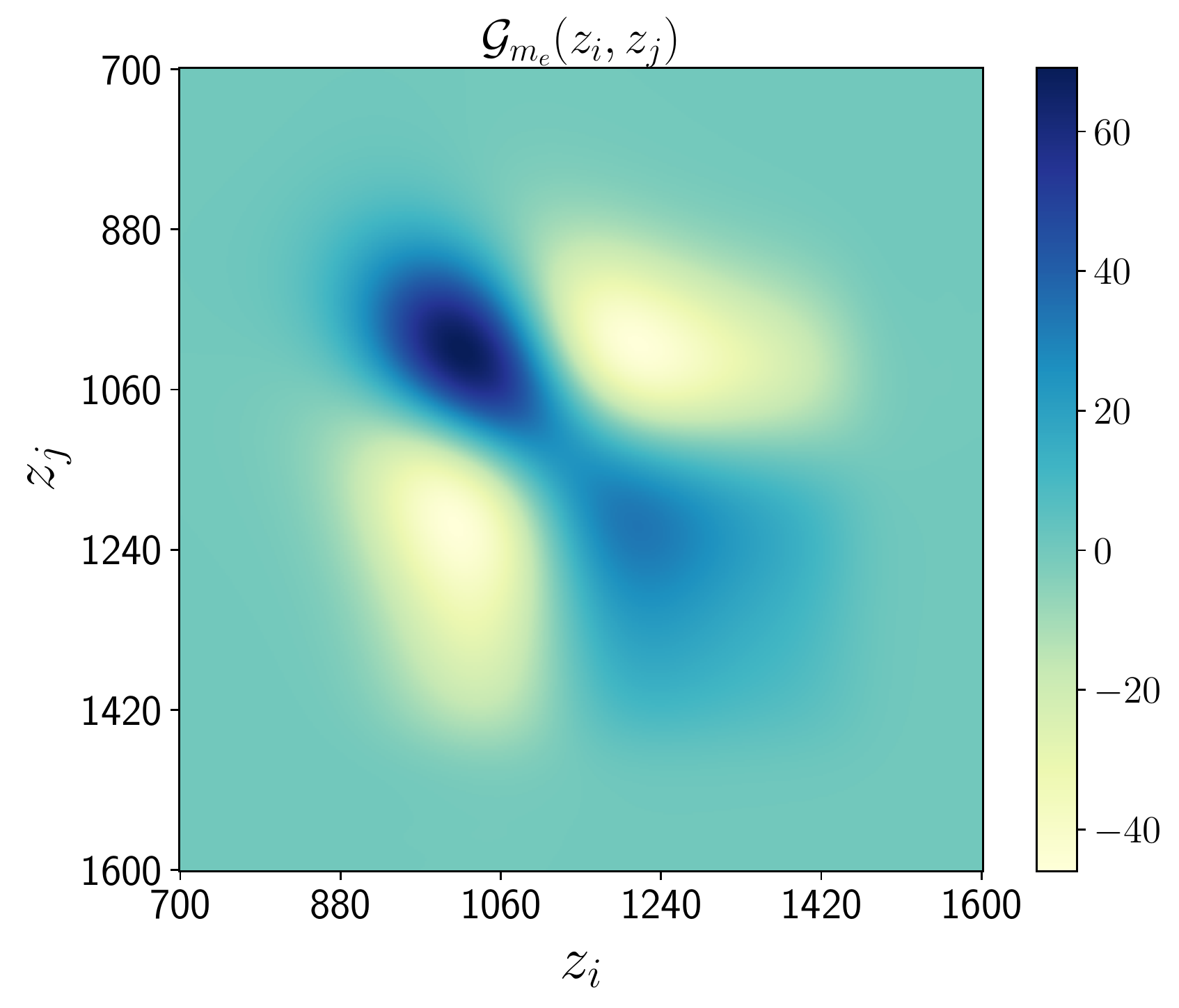}
\includegraphics[width = .32\columnwidth,trim= 10 10 00 00]{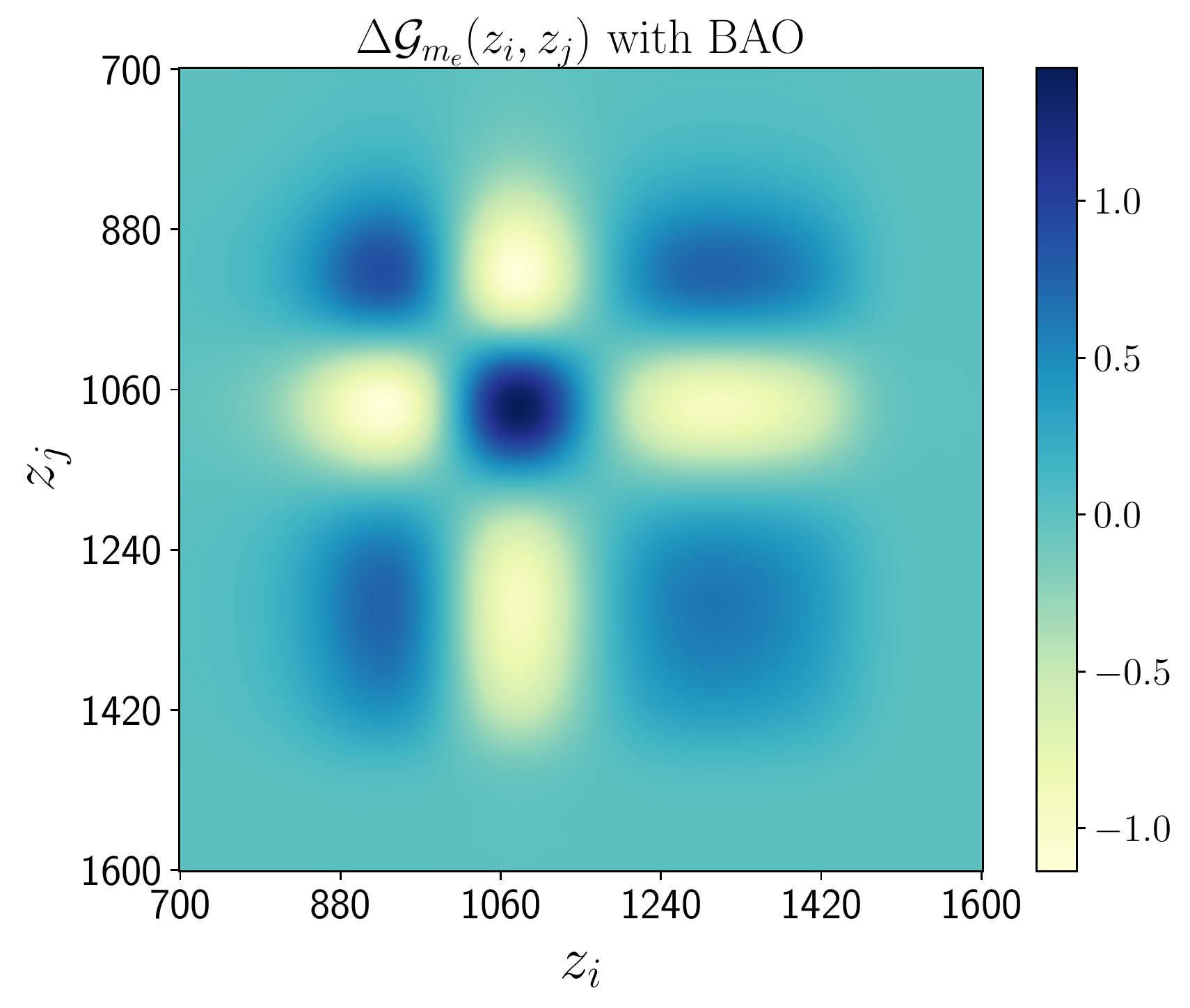}
\includegraphics[width = .32\columnwidth,trim= 10 10 00 00]{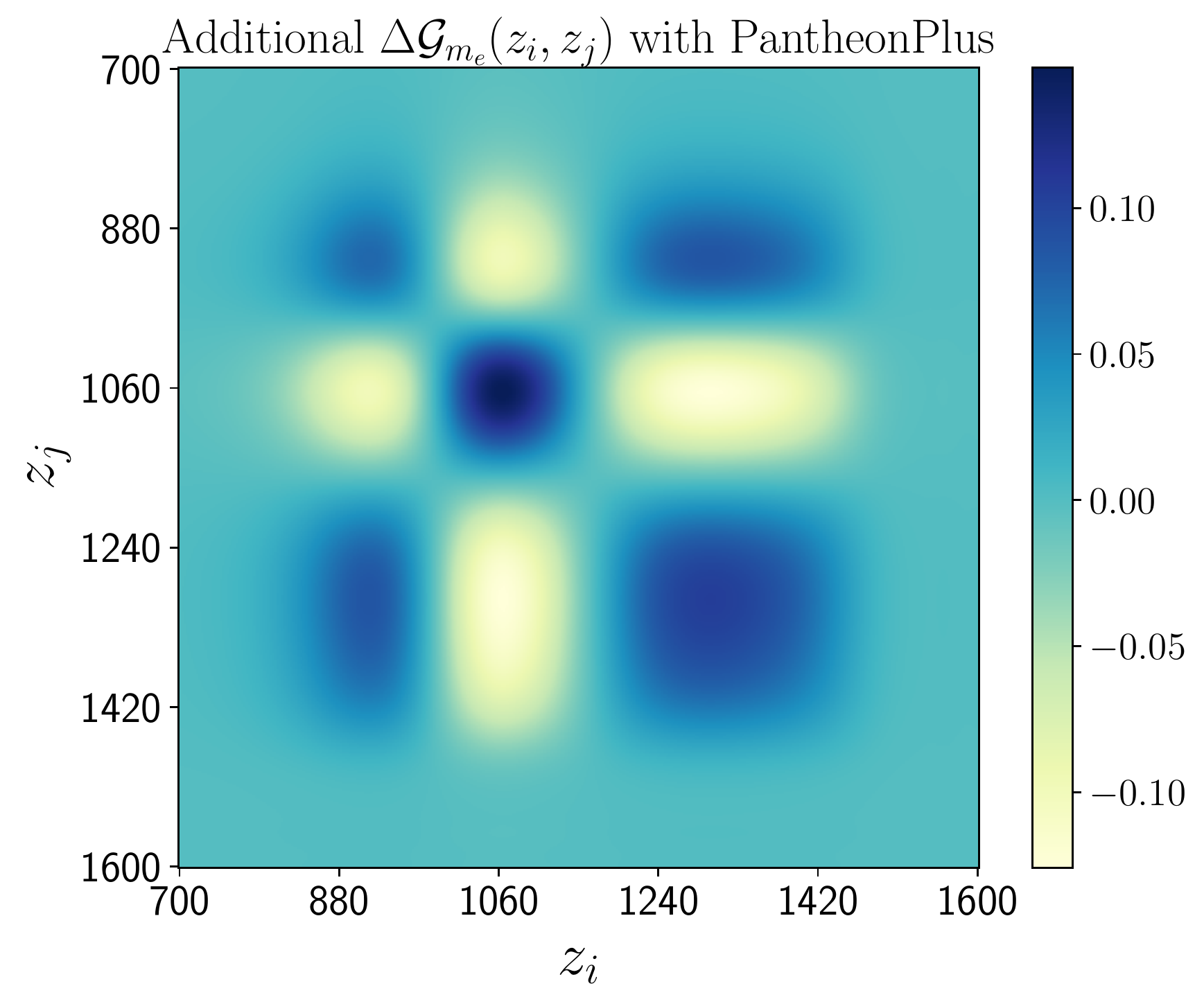}
\caption{Top: functional derivatives of best-fit chi-squared (Eq.~\eqref{eq:dchi2bf_lin}) with three data sets (Planck CMB, Planck CMB + BOSS DR12 BAO, Planck CMB + BOSS DR12 BAO + PantheonPlus). Bottom left: quadratic response of a change in best-fit chi-squared $\mathcal{G}_{m_e}(z_i,z_j) \equiv \delta^2 \chi^2_{\rm BF}/\delta \ln m_e(z_i) \delta \ln m_e (z_j)$ (Eq.~\eqref{eq:dchi2bf_quad}) with respect to logarithmic change in electron mass $m_e(z)$ at each redshift when only Planck CMB anisotropy data is considered. Bottom middle: the additional quadratic contribution when BAO is added. Bottom right: the additional quadratic contribution when PantheonPlus is added to Planck CMB + BAO.}
\label{fig:deriv-chi2}
\end{figure}

Figures~\ref{fig:deriv-Omega} and \ref{fig:deriv-chi2} shows the functional derivatives of Planck's best-fit parameters, best-fit chi-squared, and quadratic response of a change in best-fit chi-squared $\delta^2 \chi^2_{\rm BF}/\delta \ln m_e(z_i) \delta \ln m_e (z_j)$ with respect to small perturbations in $\ln m_e(z)$. These quantities form a set of building blocks in our formalism. By comparing amplitudes of the linear response (or functional derivative) $\delta \chi^2_{\rm BF}/\delta\ln m_e(z)$ in the top panel and the quadratic response in the bottom left panel of Fig.~\ref{fig:deriv-chi2}, it can be seen that a few percents level changes in $m_e(z)$ around recombination ($z\sim 1100$) can induce an order of 10\% contribution from the quadratic response (second term in Eq.~\eqref{eq:Dchi2bf_X}) to the total $\Delta \chi^2_{\rm BF}$ implying that this quadratic contribution should not be neglected.

It is useful to build some intuition about the shape of the functional derivatives of the best-fit parameters, in particular those of the reduced Hubble parameter $h \equiv H_0/(100$ km/s/Mpc), and of the total matter density parameter $\Omega_m$. CMB anisotropy data is particularly sensitive (at the 0.03\% level with Planck data \cite{Planck2018}) to the angular size of the sound horizon, $\theta_s = r_s/r_A$, where $r_s$ is the comoving size of the sound horizon, and $r_A$ is the comoving angular diameter distance, both at the recombination. In a flat universe, they are given by, respectively, 
\begin{equation}
    r_s = \int_{z_*}^\infty \frac{c_s(z)}{H(z)}\mathrm{d}z \qquad \mathrm{and} \qquad  r_A = \int_0^{z_*} \frac{1}{H(z)}\mathrm{d}z,
\end{equation}
where $z_*$ is the redshift of last-scattering of CMB photons. We may approximate the Hubble rate as follows in the $r_s$ and $r_A$ integrands, respectively:
\begin{equation}\label{eq:app:hubble}
     H(z) \approx 100~ \mathrm{km/s/Mpc} ~\sqrt{\Omega_m h^2 (1+z)^3 + \omega_r (1+z)^4} \ \ \ \ \ \textrm{and}  \ \ \ \ \ 
     H(z) \approx H_0 \sqrt{\Omega_m (1 + z)^3 + 1 - \Omega_m}.
\end{equation}
The radiation density $\omega_r$ is fixed by FIRAS measurements of the CMB monopole \cite{Mather:1993ij, Fixsen:1996nj}, and the sound speed $c_s(z)$ is weakly dependent on cosmology (and does not directly depend on $H_0$ nor $\Omega_m$). Close to the $\Lambda$CDM best-fit, one thus has
\barr
\frac{\partial \ln r_A}{\partial \ln \Omega_m}  &\approx& -0.4,~ \quad \frac{\partial \ln r_A}{\partial \ln h} \approx -1,~~~~\quad \frac{\partial \ln r_A}{\partial \ln z_*} \approx 0, \\
\frac{\partial \ln r_s}{\partial \ln \Omega_m}  &\approx& -0.23,\quad
   \frac{\partial \ln r_s}{\partial \ln h} \approx -0.48,\quad
  \frac{\partial \ln r_s}{\partial \ln z_*} \approx -0.66.
\earr
which can be combined to give
\be
\frac{\partial \ln \theta_s}{\partial \ln \Omega_m}  \approx 0.17, \quad
\frac{\partial \ln \theta_s}{\partial \ln h} \approx 0.52,\quad
\frac{\partial \ln \theta_s}{\partial \ln z_*} \approx -0.66.
\ee
\begin{figure}[H]
\centering
    \includegraphics[width=0.49\textwidth,trim= 00 20 00 20]{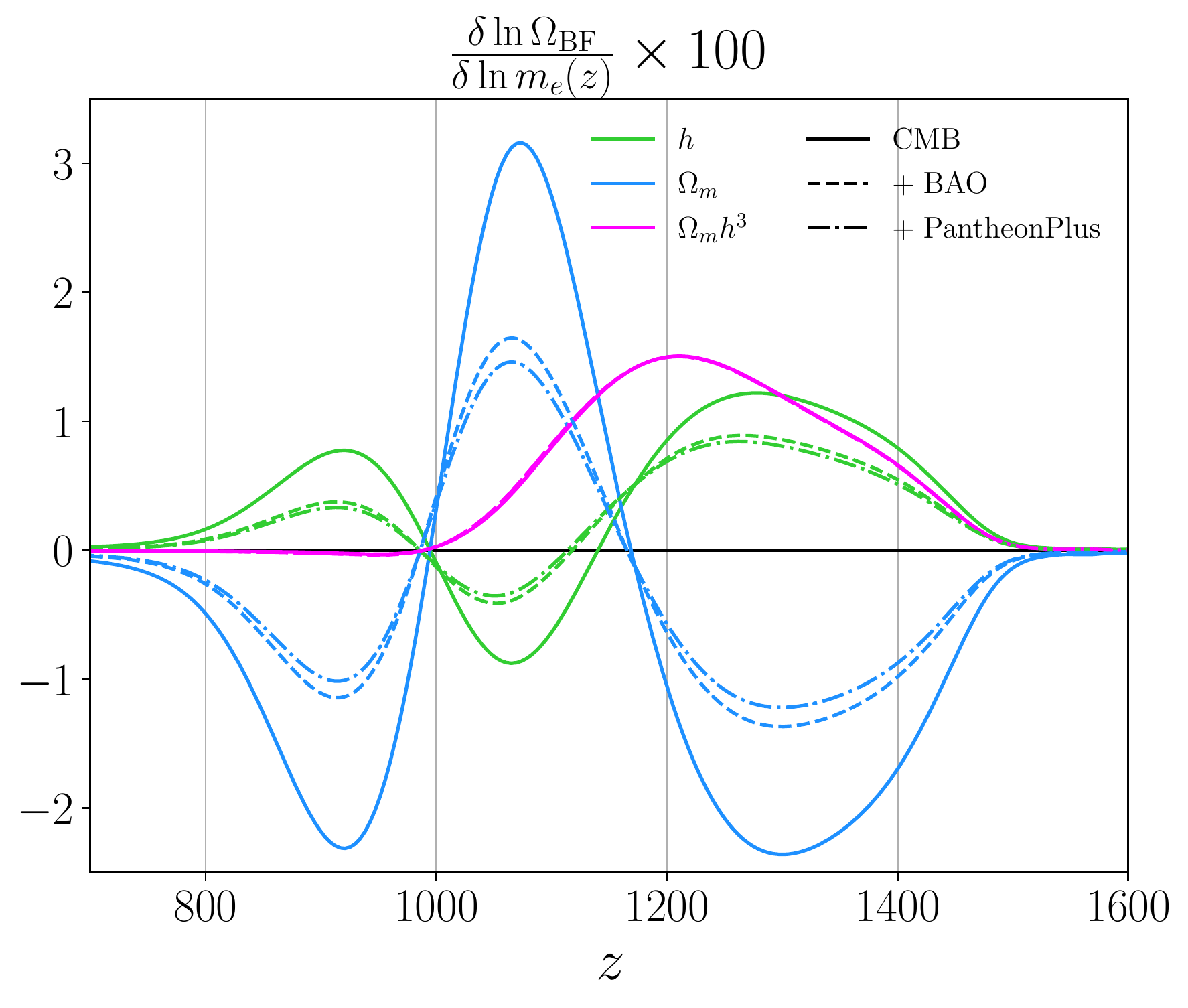}
    \caption{Functional derivatives of the best-fit parameters with respect to the fractional changes in electron mass $m_e(z)$ translated to the basis of $\{\Omega_m, \Omega_m h^3, h\}$ in order to simplify the intuition about changes of $h$. Dashed (dashdot) lines correspond to the inclusion of BAO (BAO+PantheonPlus) data.}\label{fig:parametershift}
\end{figure}

\noindent
As a consequence, we find that $\theta_s \propto \Omega_m^{0.17} h^{0.52} z_*^{-0.66} \propto (\Omega_m h^3 z_*^{-3.88})^{0.17}$ near the best-fit $\Lambda$CDM cosmology. This shows the well-known dependence of $\theta_s$ on $\Omega_m h^3$ for fixed $z_*$, which is numerically confirmed in Ref. \cite{Planck2018}. It also shows that a fractional change in the last-scattering redshift by $\Delta \ln z_*$ would affect the combination $\Omega_m h^3$ by a fractional change $\Delta \ln (\Omega_m h^3) = 3.88 \Delta \ln z_*$ in order to keep $\theta_s$ fixed. Note that this derivation does not assume anything precise about the recombination history, just that the visibility function peaks at some redshift $z_*$. 

These features are confirmed in Fig.~\ref{fig:parametershift}, which shows the functional derivatives of the best-fit $h, \Omega_m$ and $\Omega_m h^3$ with respect to $\ln m_e(z)$. We see that for $z \lesssim 1000$, the best-fit $h$ and $\Omega_m$ change in opposite ways, so as to maintain $\Omega_m h^3$ constant; this stems from the fact that changes in the electron mass in the low-redshift tail of the visibility function do not affect its peak, i.e.~do not change $z_*$. At $z \gtrsim 1000$, a positive change in $m_e$ leads to an overall speedup of recombination, i.e.~an increase in the last-scattering redshift $z_*$, hence leading to an increase in $\Omega_m h^3$ in order to keep $\theta_s$ constant (see Ref.~\cite{Cyr-Racine:2021oal,Ge:2022qws} for a similar parameter degeneracy induced by uniform rescaling of the various rates involved in the Boltzmann equations which are driving the CMB physics).

An important result visible in Fig.~\ref{fig:parametershift} is that at almost any point in the thermal history, an upward shift in best-fit $h$ results in a corresponding downward shift in best-fit $\Omega_m$. This leads us to conclude that most simple smooth solutions to the Hubble tension should lead to a downward shift in $\Omega_m$ and correspondingly cause issues with late-time probes such as supernovae (e.g. Pantheon+, $\Omega_m = 0.334\pm0.018$ \cite{Brout:2022vxf}), BAO (e.g. from BOSS \cite{eBOSS:2020yzd}, $\Omega_m = 0.298 \pm 0.016$), and possibly other late-time observables (such as cosmic chronometers).  The strength of this inconsistency is only limited by the experimental precision of these late-time probes, which is expected to strongly increase in the near future (e.g. with DESI \cite{DESI:2016fyo}, Euclid \cite{Refregier:2010ss}, LSST \cite{LSSTScience:2009jmu}, ...), possibly ruling out solutions to the Hubble tension based on shifts in recombination. It should be mentioned, however, that currently still loopholes exist, caused by increasing the freedom in the late-time expansion history. Indeed, in \cite{Schoneberg:2021qvd} the model of a varying electron mass was supplemented by curvature in order to weaken the strong constraining power of the BAO and supernovae, leading to still an overall good fit. We stress, however, that future data are likely to better constrain the late time expansion history, eliminating this loophole. Hope then remains only for relatively complicated solutions that precariously balance upward and downward shifts of $\Omega_m$ and $H_0$ coming from the different regions in such a way that overall $\Omega_m$ is unchanged and $H_0$ is increased.

It should also be mentioned that the inclusion of BAO (or BAO + PantheonPlus) data directly into the formalism confirms this picture (as seen in Fig.~\ref{fig:parametershift}). While the overall change in $\Omega_m h^3$ remains the same (in order to compensate for possible shifts in $\theta_s$), the individual variations in $h$ and $\Omega_m$ are much less pronounced, since the $\Omega_m$ is now more constrained by additional data. Thus, the bestfit is less likely to move into a direction of $\Omega_m$ less in agreement with such data, hence leading to smaller functional derivatives.

\section{Comparison with PCAs}\label{appendix:PCAs}

\begin{figure}[ht]
\centering
    \includegraphics[width=0.50\textwidth,trim= 00 17 00 20]{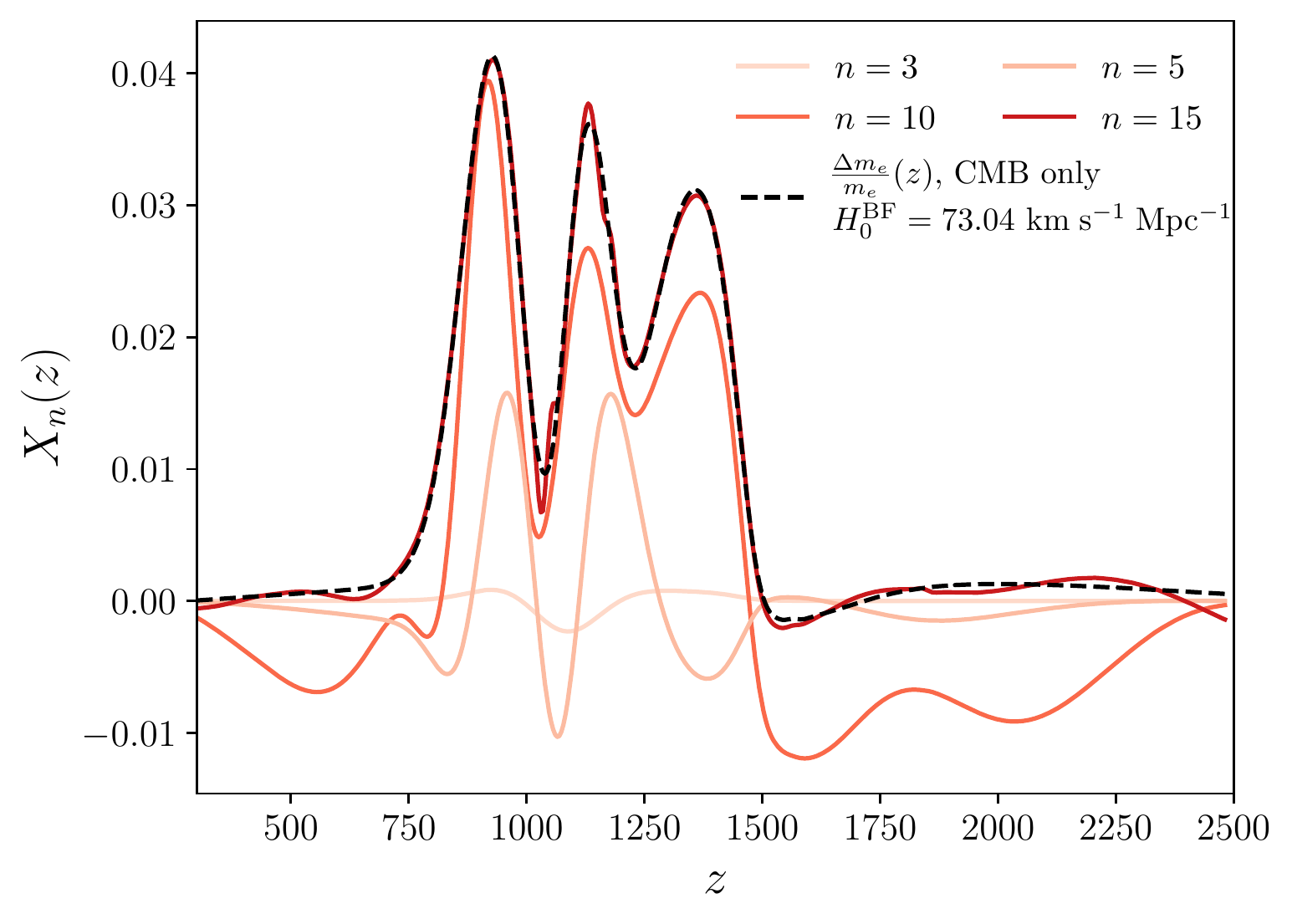}
    \includegraphics[width=0.48\textwidth,trim= 00 20 00 20]{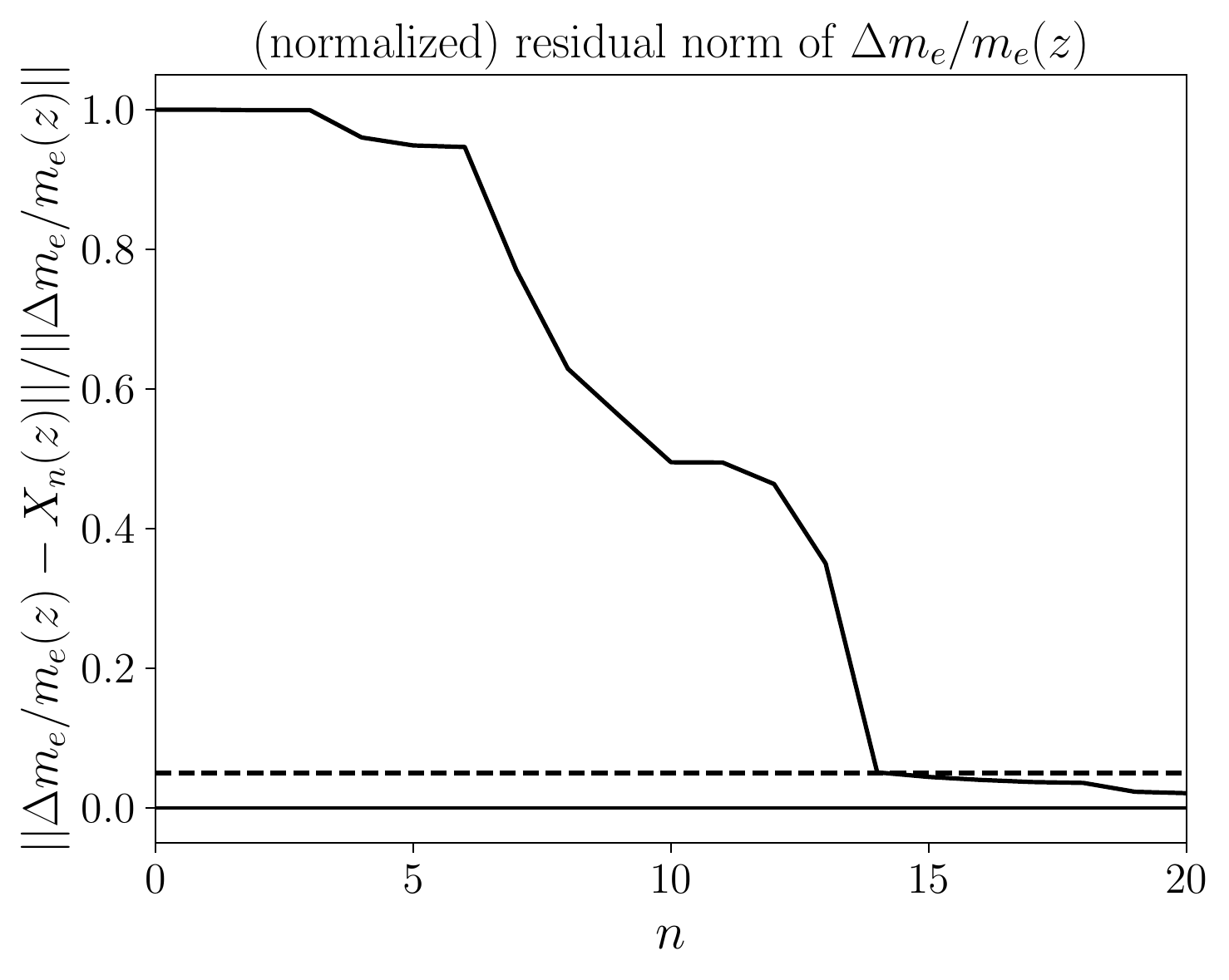}
    \caption{Left: the projected solution onto the $n$ eigenmodes with the $n$ largest eigenvalues, $X_n(z)$. Right: the residual norm of the solution (black curve in Fig.~\ref{fig:me-cmb}, also shown in the left pannel as a black dashed line) after subtracting its projections onto eigenmodes with the largest eigenvalues: $||\Delta m_e/m_e(z) - X_n(z)||/||\Delta m_e/m_e(z)||$ where $||f(x)||$ is the L2 norm of $f(x)$. The dashed line indicates 5\% residual norm.}\label{fig:residual_norm}
\end{figure}
In this appendix we make contact with Principal Component Analyses (PCAs). While analyses based on PCA primarily investigate the first few eigenmodes describing perturbations to recombination to which the data is most sensitive (e.g. Refs.~\cite{Hart:2019gvj,Hart:2021kad}), our goal is to find the smallest perturbations allowed by the data, producing a desired shift in best-fit cosmological parameters while not increasing the best-fit $\chi^2$. Consequently, in general, the solution we obtain will project onto many eigenmodes simultaneously, even those with small eigenvalues to which the data is not very sensitive. The left pannel of Fig.~\ref{fig:residual_norm} shows the residual norm of the solution (black curve in Fig.~\ref{fig:me-cmb}, also shown in the left pannel as a black dashed line) after subtracting its projections onto eigenmodes with the largest eigenvalues of the Fisher matrix marginalized over cosmological parameters [or simply Eq.~\eqref{eq:dchi2bf_quad}]. We checked that the first three eigenmodes agree with those of Ref.~\cite{Hart:2021kad}. One can clearly see that the solution we obtain cannot be characterized by just a few principal components. For example, to achieve the residual norm to be less than 5\%, at least 15 eigenmodes are needed. The fact that the first three eigenmodes do not significantly contribute to our solution agrees with the results of Ref.~\cite{Hart:2021kad}: the first three principal components for $m_e$ do not provide a large enough basis for promising solutions to the Hubble tension (the same holds for $\alpha$).

\begin{figure}[ht]
\centering
\includegraphics[width = .9\columnwidth,trim= 00 30 00 00]{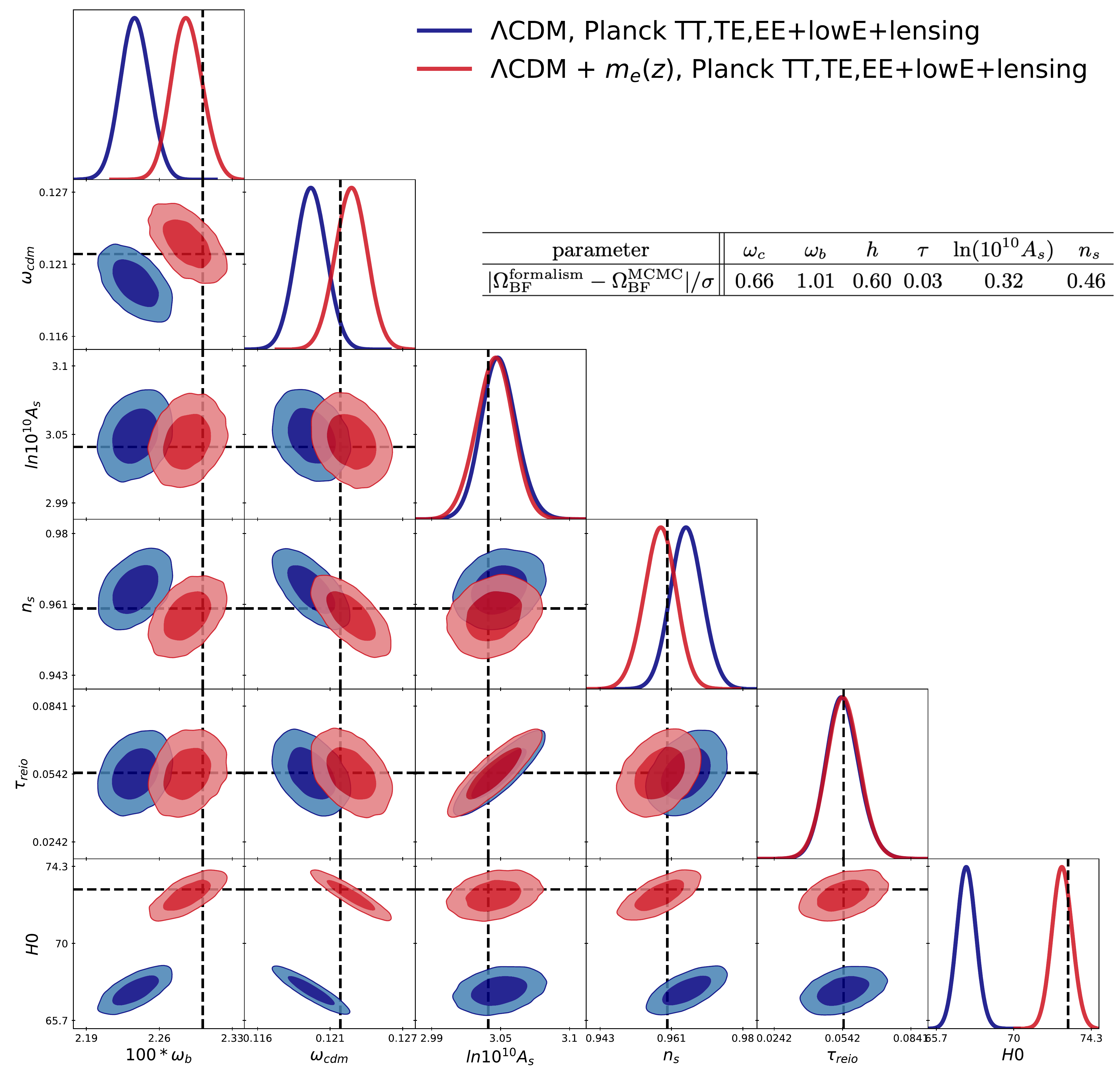}
\caption{Contour plot with two models, $\Lambda$CDM and $\Lambda$CDM + $m_e(z)$. The latter model is what's found with Planck CMB data as a solution to the Hubble tension with SH0ES, which is shown as black curve in Fig.~\ref{fig:me-cmb}. The black dashed lines are the estimated new best-fits by our formalism, Eq.~\eqref{eq:DObf_X}. Biases in terms of uncertainty of each parameter are shown in the inset table. Note that the $\omega_b$ in the $\Lambda$CDM + $\alpha(z)$ model becomes less consistent with the BBN constraint from Ref.~\cite{Pisanti:2020efz} ($\omega_b = 0.0220 \pm 0.0005$) obtained using the helium abundance measurements from Refs.~\cite{Aver:2015iza,Peimbert:2016bdg,Hsyu:2020uqb}, but more consistent with those derived using the Helium mass fraction found by Ref.~\cite{Izotov:2014fga} ($\omega_b = 0.0234 \pm 0.0005$). Note that, however, while the constraints on the number of relativistic species $N_{\rm eff}$ \cite{Pisanti:2020efz} based on Ref.~\cite{Aver:2015iza,Peimbert:2016bdg,Hsyu:2020uqb} are consistent with the fixed value $N_{\rm eff}=3.046$ in our analysis, the constraint based on Ref.~\cite{Izotov:2014fga}, $N_{\rm eff}=3.60\pm0.17$, is not consistent with it.}
\label{fig:MCMC-me}
\end{figure}

\section{Validation test - time-varying electron mass $m_e(z)$ from Planck CMB}
\label{appendix:validation}

Here we show a validation test for our formalism. Fig.~\ref{fig:MCMC-me} shows the MCMC results of the $\Lambda$CDM + $m_e(z)$ model, where $m_e(z)$ is found by our formalism with Planck CMB only to be a solution to the Hubble tension (black curve of Fig.~\ref{fig:me-cmb}). The black dashed lines are the estimated new best-fits by the formalism, Eq.~\eqref{eq:DObf_X}. $\Lambda$CDM contours are also given to better illustrate how much shifts occured in best-fit parameters. As shown in the inset table, the inconsistencies (biases) of each parameter between our formalism and MCMC results are all within $\sim1\sigma$. These small inconsistencies are simply due to the fact that our formalism is not exact. Note, moreover, that small biases in $A_s$ and $\tau$ are expected from using the compressed low-$\ell$ likelihood from Ref.~\cite{Prince:2021fdv}.

\section{Time-varying electron mass $m_e(z)$ from Planck + BAO or Planck + BAO + PantheonPlus}

\begin{figure}[H]
\centering
\includegraphics[width = .45\columnwidth,trim= 05 20 05 10]{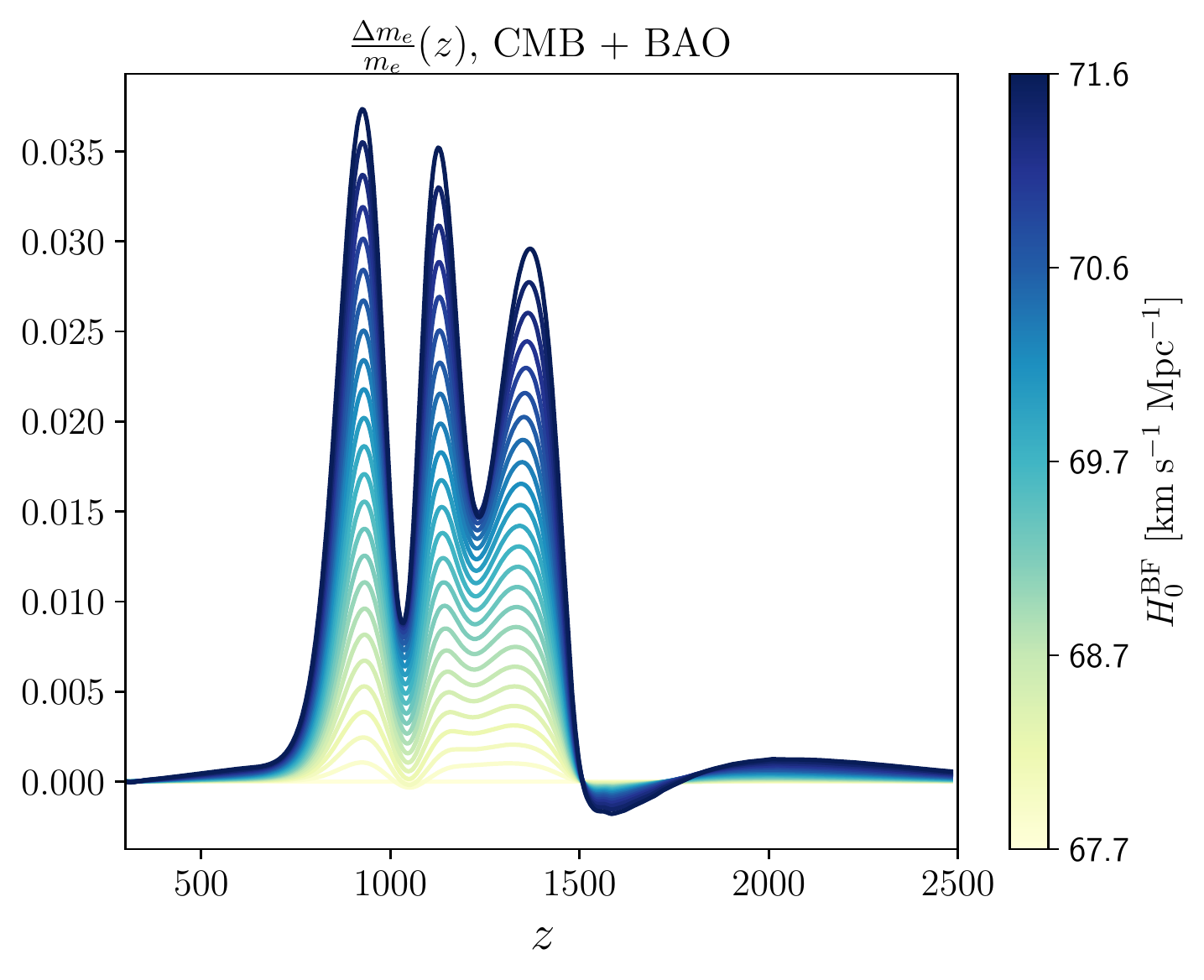}
\includegraphics[width = .45\columnwidth,trim= 05 20 05 10]{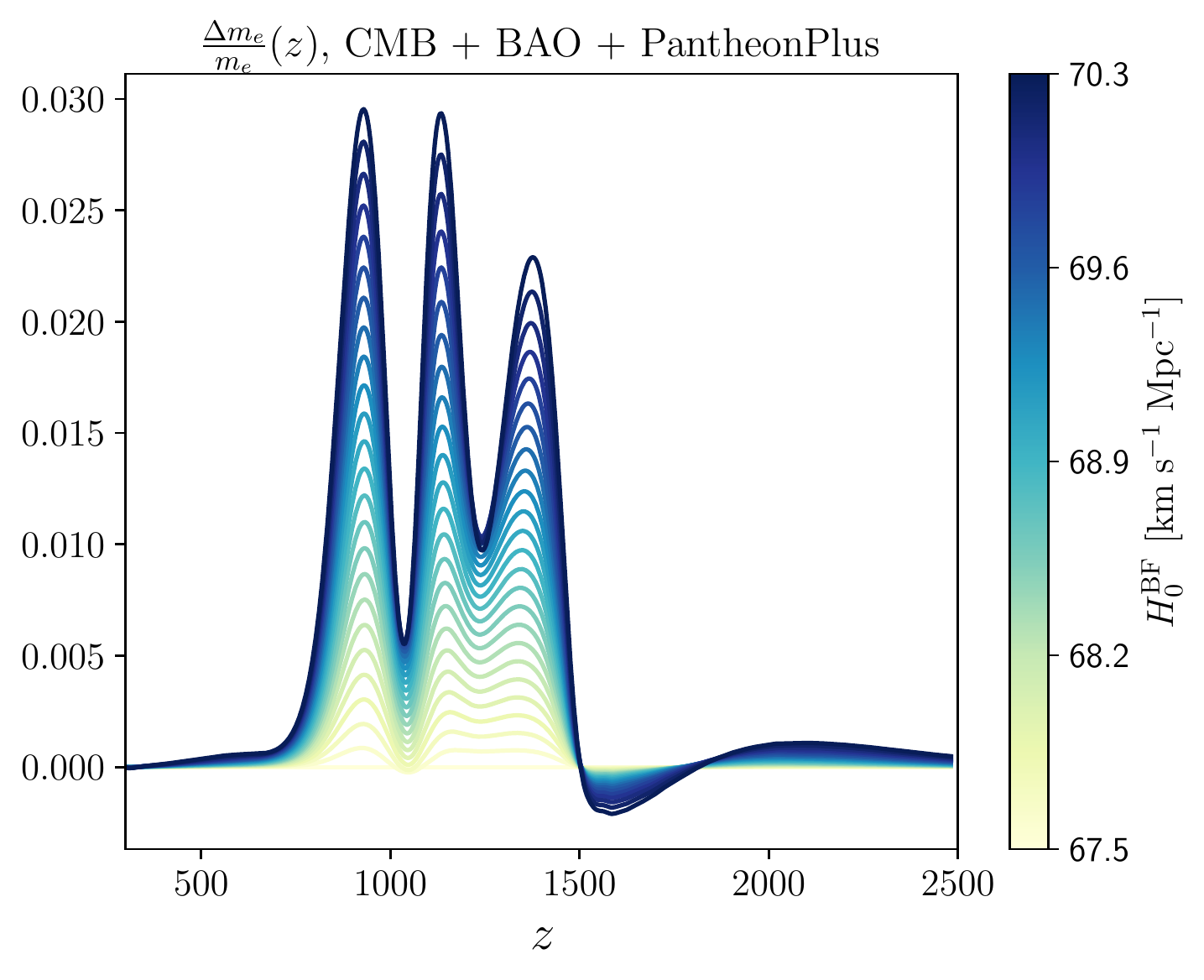}
\caption{Solutions for $\frac{\Delta m_e}{m_e}(z)$ given a required shift in the CMB + BAO (left) and CMB + BAO + PantheonPlus (right) best-fit Hubble constant $H_0$, using Planck data \cite{Planck2018} for CMB, BOSS DR12 anisotropic measurements \cite{BOSS:2016wmc} for BAO, the constraint on $\Omega_m$ from PantheonPlus \cite{Brout:2022vxf}. The upper bound of $H_0^{\rm BF}$ is determined as the maximum value of $H_0^{\rm BF}$ which gives a self-consistent solution satisfying $\Delta \chi^2<1$ and $|\Omega_{\rm BF}^{\rm formalism} - \Omega_{\rm BF}^{\rm MCMC}|/\sigma<1$ for all six cosmological parameter $\Omega$'s.}
\label{fig:me-bao-pantheon}
\end{figure}

As we add more data (BAO and PantheonPlus), the constraints on comsmological parameters get tighter which makes the approximation of Taylor-expansion of $\chi^2$ break down more easily with shifts of best-fit cosmology (see Appendix.~\ref{appendix:error-approximations}), hence it becomes more difficult to achieve large $H_0^{\rm BF}$ values with self-consistent solution. See Fig.~\ref{fig:me-bao-pantheon} for solutions for $\frac{\Delta m_e}{m_e}(z)$ with two different data sets used, Planck + BAO and Planck + BAO + PantheonPlus.

\section{Estimating errors from the approximations taken in the formalism}
\label{appendix:error-approximations}

While the approximations taken in our formalism make the optimization problem tractable, this in principle makes our results not-exact. In this Appendix, we show how large the errors induced by these approximations can be, by comparing Planck CMB Gaussian $\chi^2$'s in our final results with Planck CMB anisotropy data. First, we define the reference chi-squared $\chi^2_{\rm ref}$, which is corresponding to a new best-fit chi-squared estimated by our formalism with a solution (perturbation in $f$) $\Delta f(z)$ which is obtained with a target $H_0^{\rm BF}=73.04\;\text{km\;s}^{-1}\text{Mpc}^{-1}$. Note that this chi-squared is approximately the same as the Planck $\Lambda$CDM best-fit chi-squared, i.e. $\chi^2_{\rm ref} = \chi^2_{\Lambda{\rm CDM},{\rm BF}}$, since the obtained $\Delta f(z)$ during the minimization process satisfies $\Delta \chi^2 =0$. We define two more effective chi-squares which are obtained by lifting either the approximation of Taylor-expansion of $\chi^2$ in the cosmological parameters or the assumed linearity of the $C_\ell$'s in $\Delta f(z)$,
\barr
\chi^2_{\rm non-Taylor} &\equiv& \chi^2 \Big\{C_\ell(\vec{\Omega}_{\rm BF}^f) +  \Delta C_\ell[\Delta f (z)]\Big\}, \nonumber\\
\chi^2_{\rm non-lin} &\equiv& \chi^2 \Big \{C_\ell(\vec{\Omega}_{\rm fid};\;\Delta f(z)) + \Delta C_\ell[\vec{\Omega}_{\rm BF}^f - \vec{\Omega}_{\rm fid}]\Big\},
\earr
where $\vec{\Omega}_{\rm fid}$ is the fiducial cosmology, $\vec{\Omega}_{\rm BF}^f$ is estimated best-fit cosmology from the formalism using Eq.~\eqref{eq:DObf_X} for a given $\Delta f(z)$, and
\be
 \Delta C_\ell[\Delta  f (z)] \equiv \int dz\; \frac{\delta C_\ell}{\delta f(z)} \Delta f(z),\quad
 \Delta C_\ell[\vec{\Omega}^f_{\rm BF} - \vec{\Omega}_{\rm fid}] \equiv \frac{\delta C_\ell}{\delta \Omega^i} (\vec{\Omega}^f_{\rm BF} - \vec{\Omega}_{\rm fid}).
\ee
That is, we are comparing the $\chi^2_\mathrm{ref}$ of our formalism that involves both the Taylor-expansion and the linearization in $\Delta f$ to the $\chi^2$ values we would obtain by dropping either approximation. This can give us an estimate of how impactful each of the two approximations is.
In addition to time-varying electron mass $m_e(z)$ and fine structure constant $\alpha(z)$ which are considered in this work, we additionally consider the net recombination rate $R(z)$, i.e.~the right-hand-side of $\frac{dx_e}{d\ln a}=-R(x_e,a)$, where $x_e$ is the free electron fraction and $a$ is the scale factor.

The estimated error due to each approximation for all three cases is shown in Table~\ref{tab:error-approximations}. In cases of $f(z)=\delta \ln m_e(z),~\delta\ln \alpha(z)$, errors from two approximations are comparable and these are responsible for resulting biases shown in Appendix.~\ref{appendix:validation} (Note that also the assumed Gaussianity in the data could be responsible for biases as well). When $f(z)=\delta \ln R(z)$, errors from two approximations are larger and this is the reason why we do not consider it in the main text as an extension for a solution to the Hubble tension.
\begin{table}[H]
  \centering
  \begin{tabular}{c||cc|cc}
  \hline
   &  ~~$\chi^2_{\rm non-Taylor}$~~ &  $\frac{|\chi^2_{\rm non-Taylor}-\chi^2_{\rm ref}|}{\chi^2_{\rm ref}}\times 100 \%$ & ~~$\chi^2_{\rm non-lin}$~~ &  $\frac{|\chi^2_{\rm non-lin}-\chi^2_{\rm ref}|}{\chi^2_{\rm ref}}\times 100 \%$\\
    \hline
    $f(z)=\delta \ln m_e(z)$  &   677.7 & 14.2~\% &666.2 & 12.3~\% \\
    $f(z)=\delta \ln \alpha(z)$  &  683.5 & 15.2~\% & 663.2 & 11.8~\% \\
     $f(z)=\delta \ln R(z)$ &  718.1 & 21.0~\% & 774.3 & 30.5~\%\\
  \hline 
  \end{tabular}
  \caption{Note that $\chi^2_{\rm ref}=\chi^2_{\Lambda\text{CDM}, \text{BF}} =593.3$. The solutions for $f(z)$ are obtained with a target $H_0^{\rm BF}=73.04\;\text{km\;s}^{-1}\text{Mpc}^{-1}$.}
\label{tab:error-approximations}
\end{table}

\vspace{-1.5\baselineskip}
\section{Time-varying fine structure constant $\alpha(z)$}
\label{appendix:me}

\begin{figure}[ht]
\centering
\includegraphics[width = .9\columnwidth,trim= 00 20 00 00]{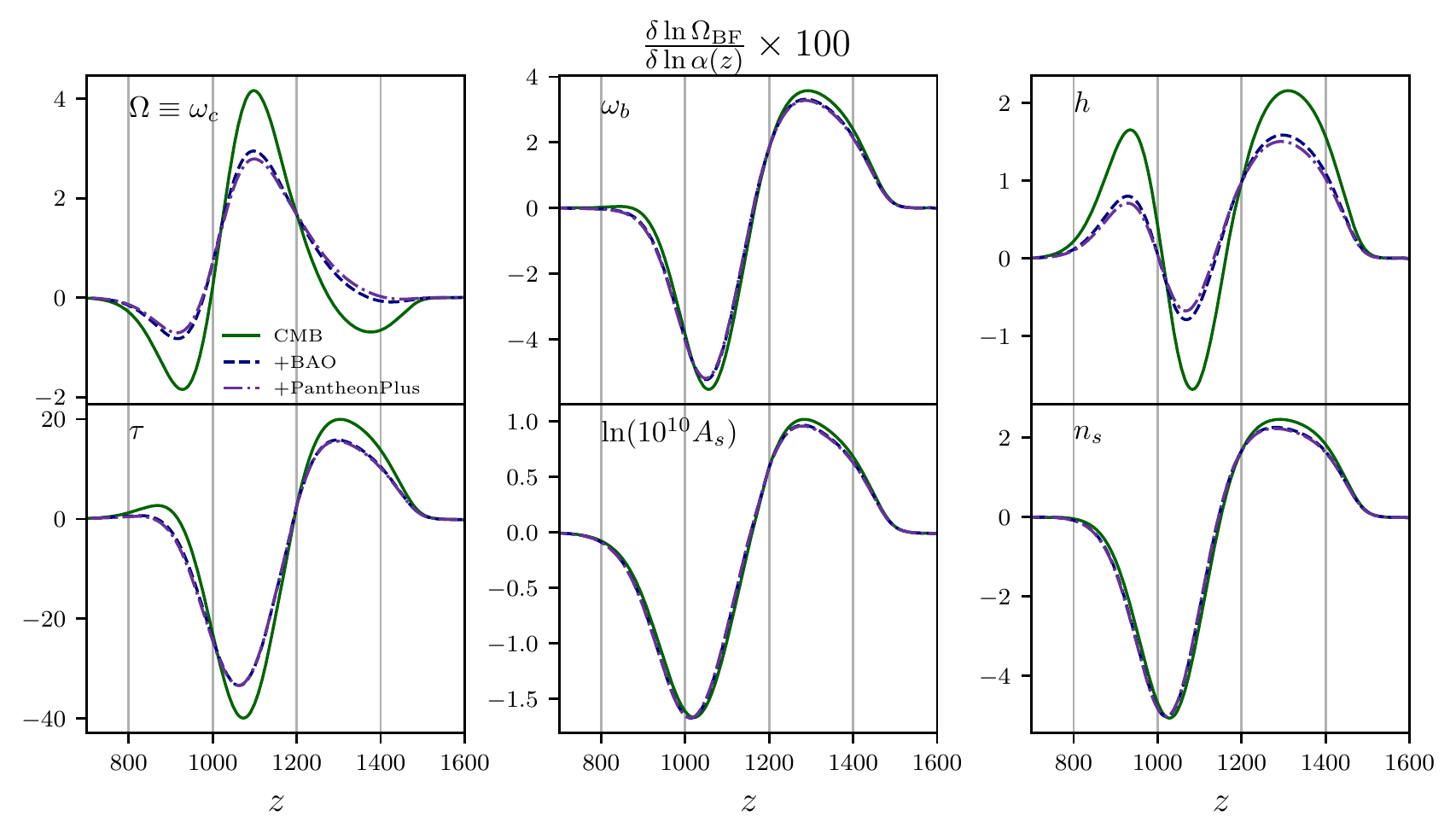}
\caption{The same plots as Fig.~\ref{fig:deriv-Omega} but with $f(z)=\ln \alpha(z)$ instead of $\ln m_e(z)$.}
\label{fig:deriv-Omega-alpha}
\end{figure}

\begin{figure}[ht]
\centering
\includegraphics[width = .4\columnwidth,trim= 10 10 00 20]{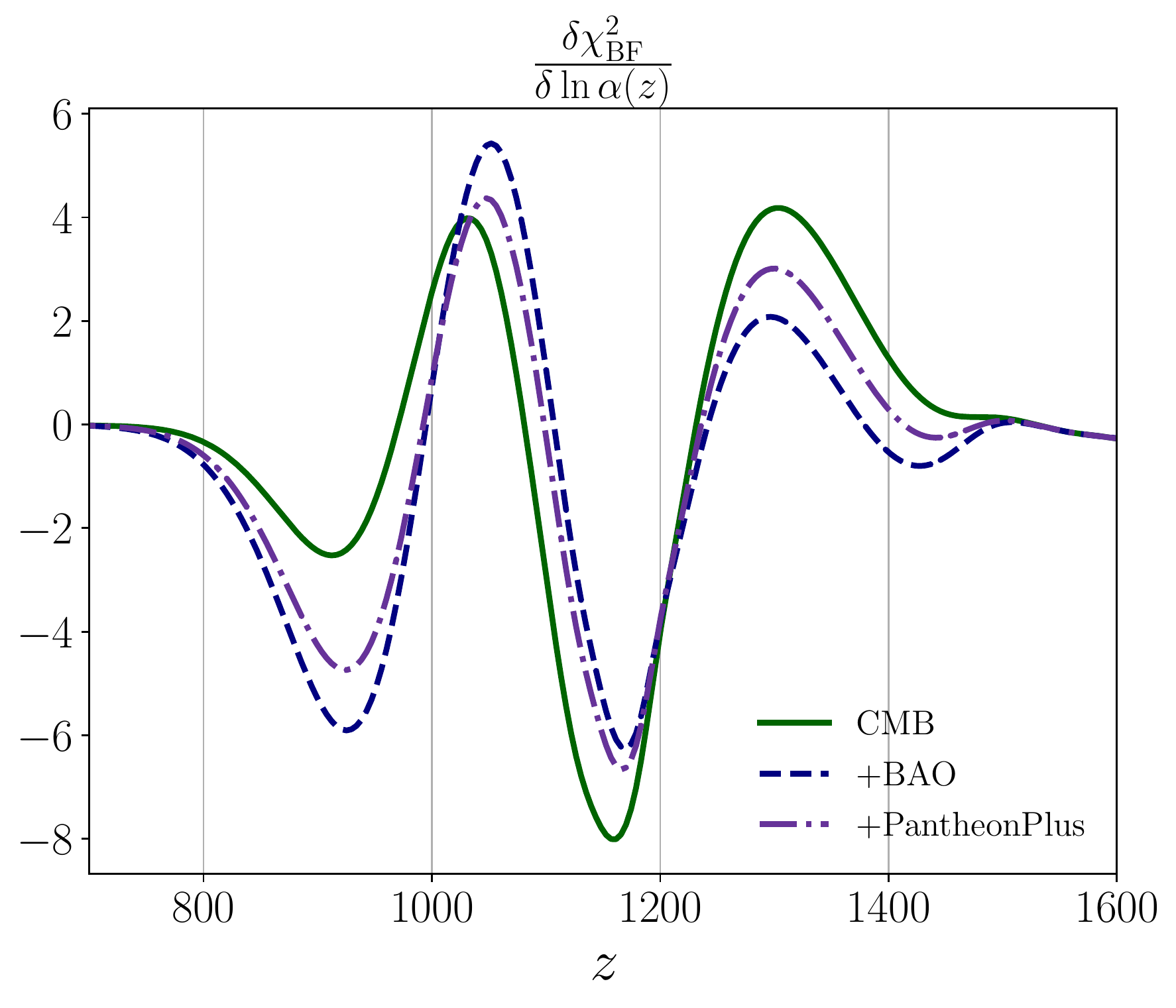}
\\\includegraphics[width = .335\columnwidth,trim= 10 30 00 00]{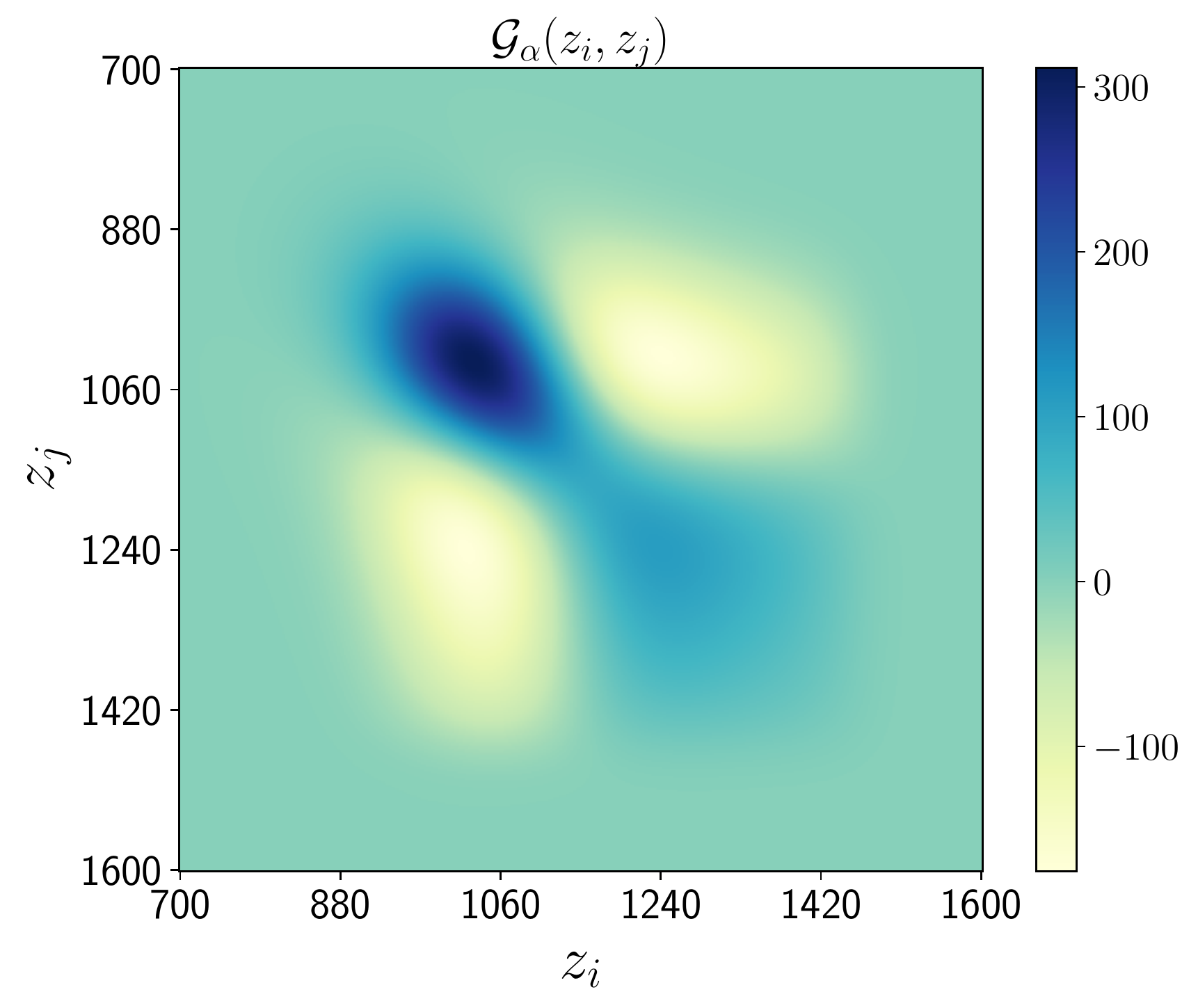}
\includegraphics[width = .32\columnwidth,trim= 10 30 00 00]{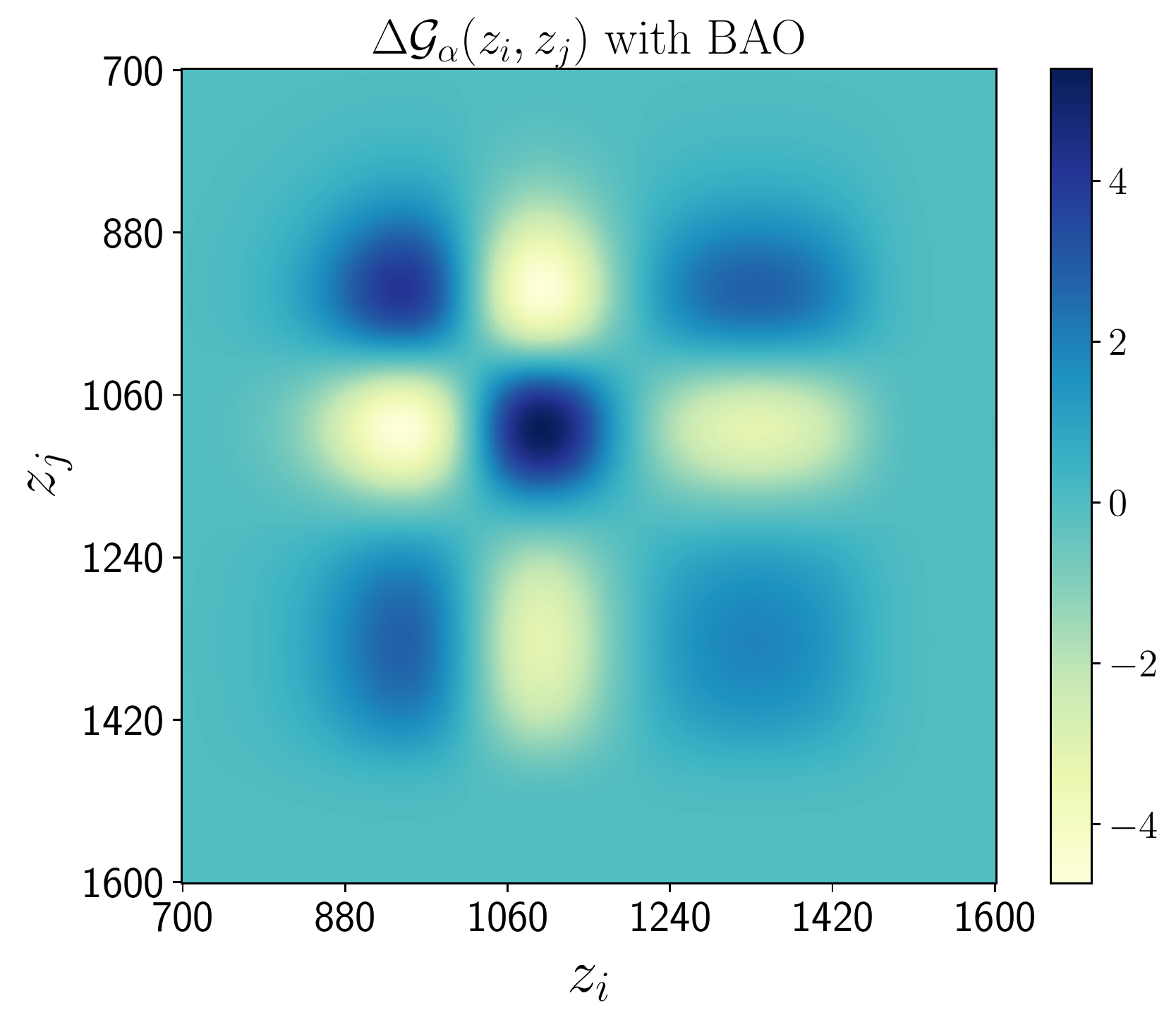}
\includegraphics[width = .325\columnwidth,trim= 10 30 00 00]{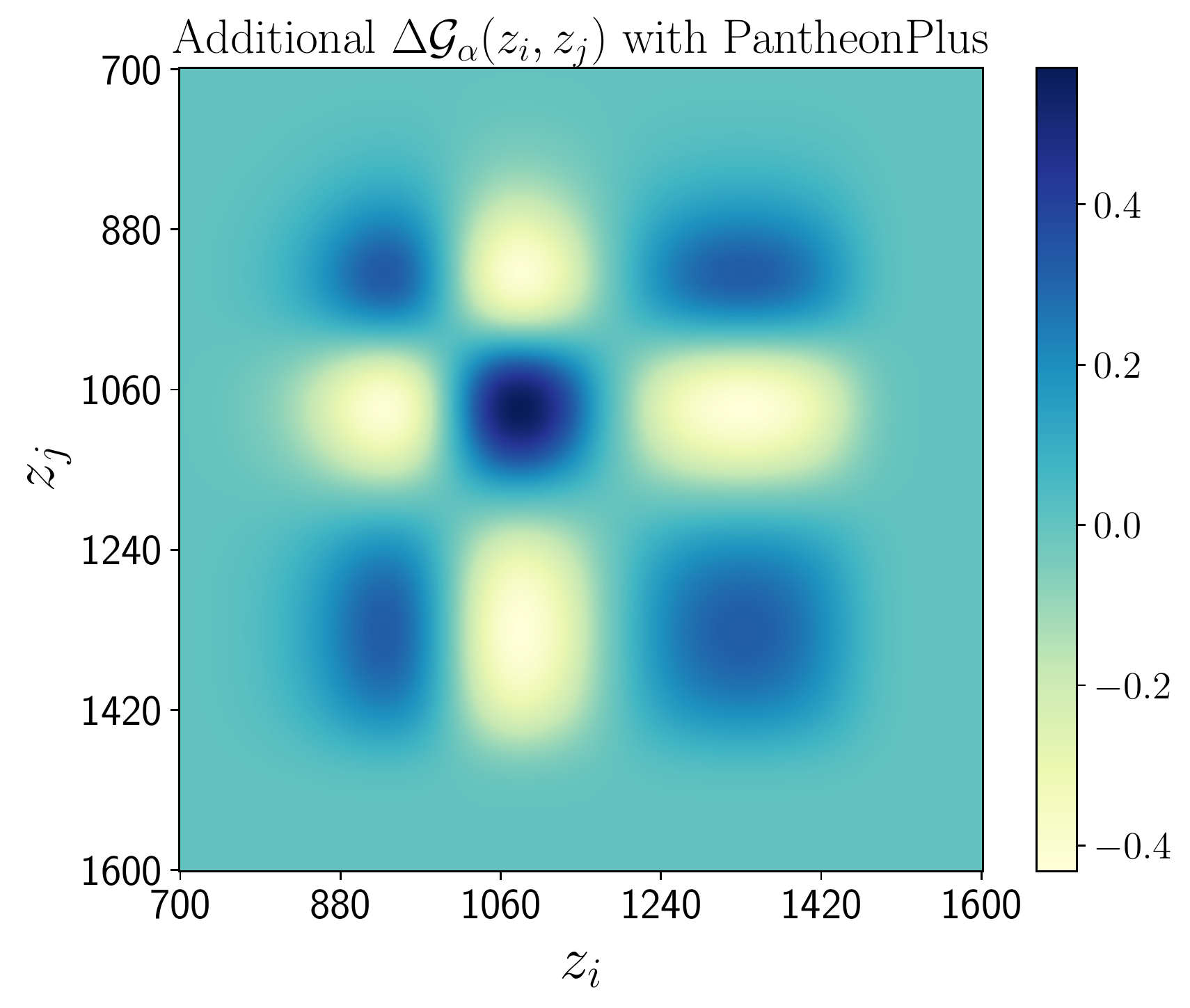}
\caption{The same plots as Fig.~\ref{fig:deriv-chi2} but with $f(z)=\ln \alpha(z)$ instead of $\ln m_e(z)$.}
\label{fig:deriv-chi2-alpha}
\end{figure}

\begin{figure}[hb]
\includegraphics[width = .325\columnwidth,trim= 07 20 07 10]{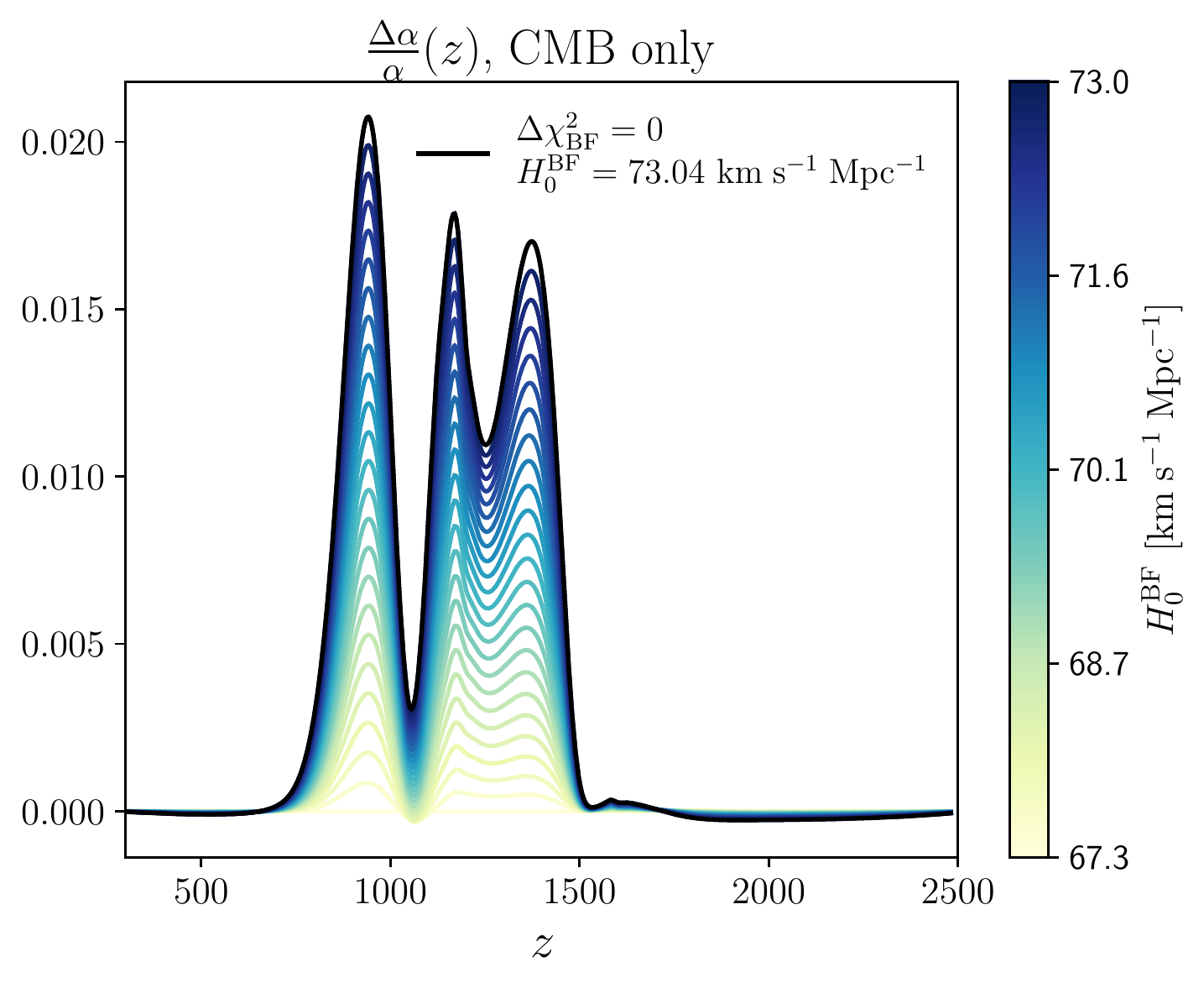}
\includegraphics[width = .33\columnwidth,trim= 07 20 06 10]{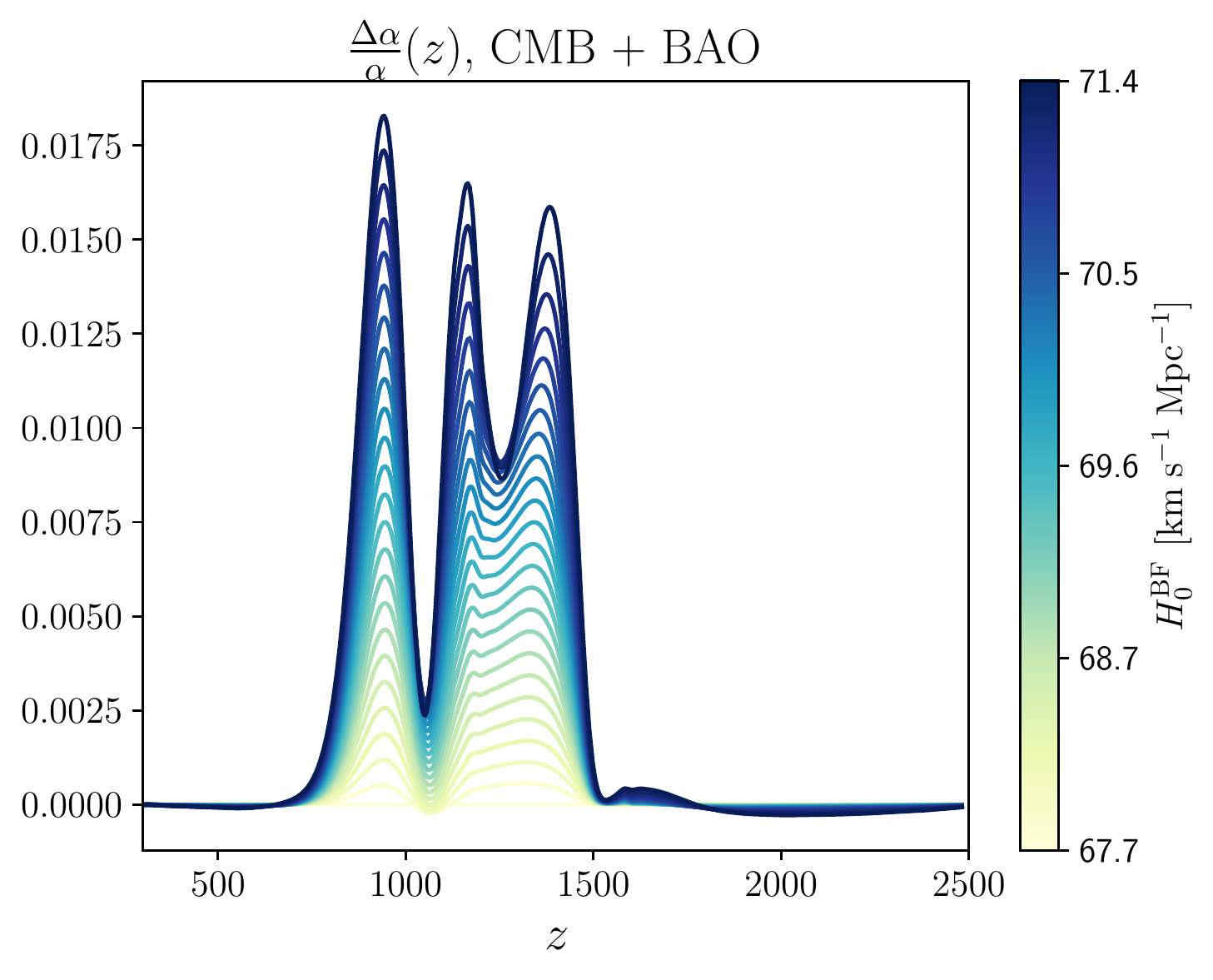}
\includegraphics[width = .325\columnwidth,trim= 07 20 07 10]{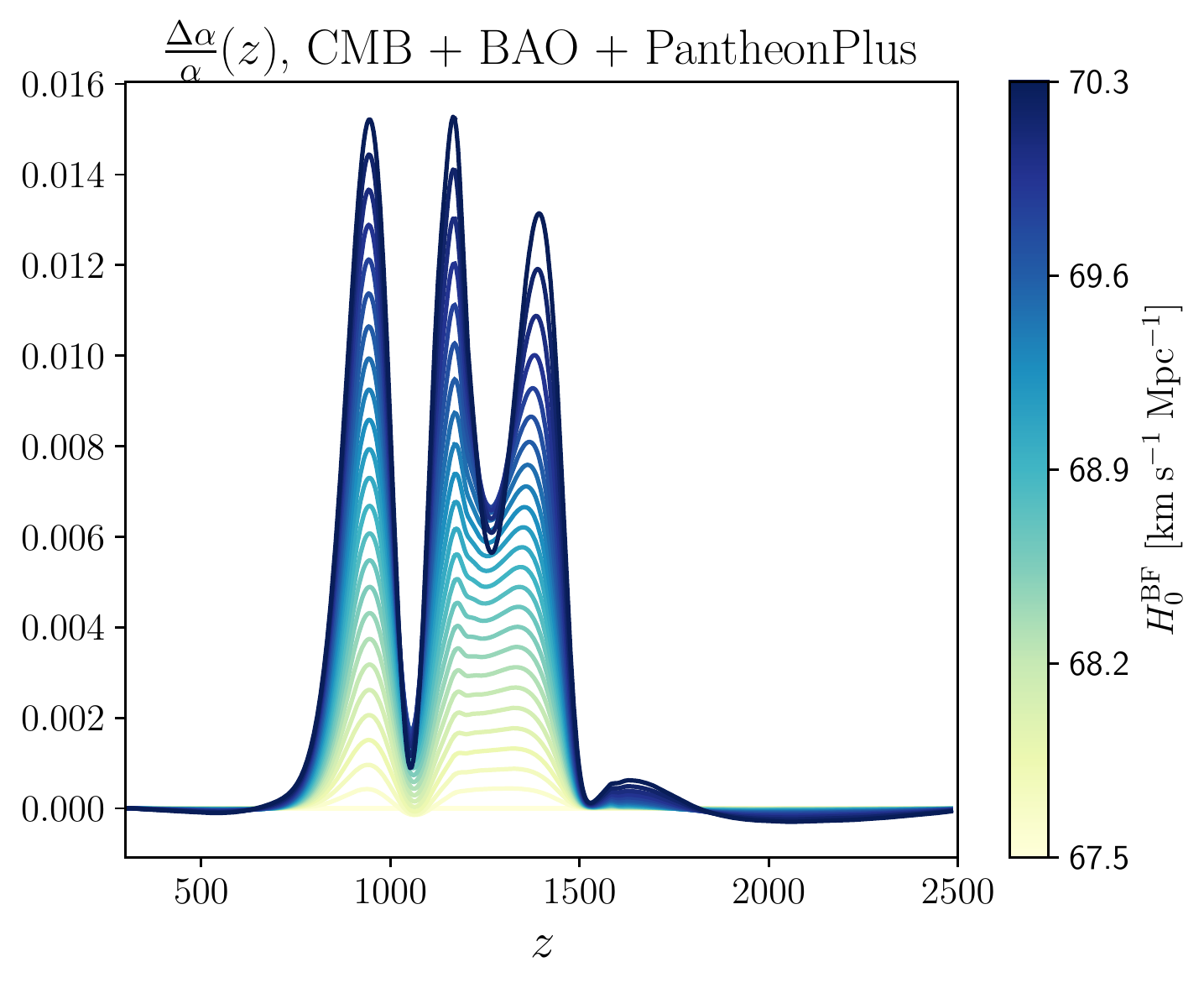}
\caption{The same plots as Fig.~\ref{fig:me-cmb} and \ref{fig:me-bao-pantheon} but with $f(z)=\ln \alpha(z)$ instead of $\ln m_e(z)$. With CMB + BAO (CMB + BAO + PantheonPlus) data, we could lower the Hubble tension down to $\sim 1.4\sigma$ ($2.4\sigma$).}
\label{fig:alpha}
\end{figure}

\begin{figure}[ht]
\includegraphics[width = .315\columnwidth,trim= 00 20 00 30]{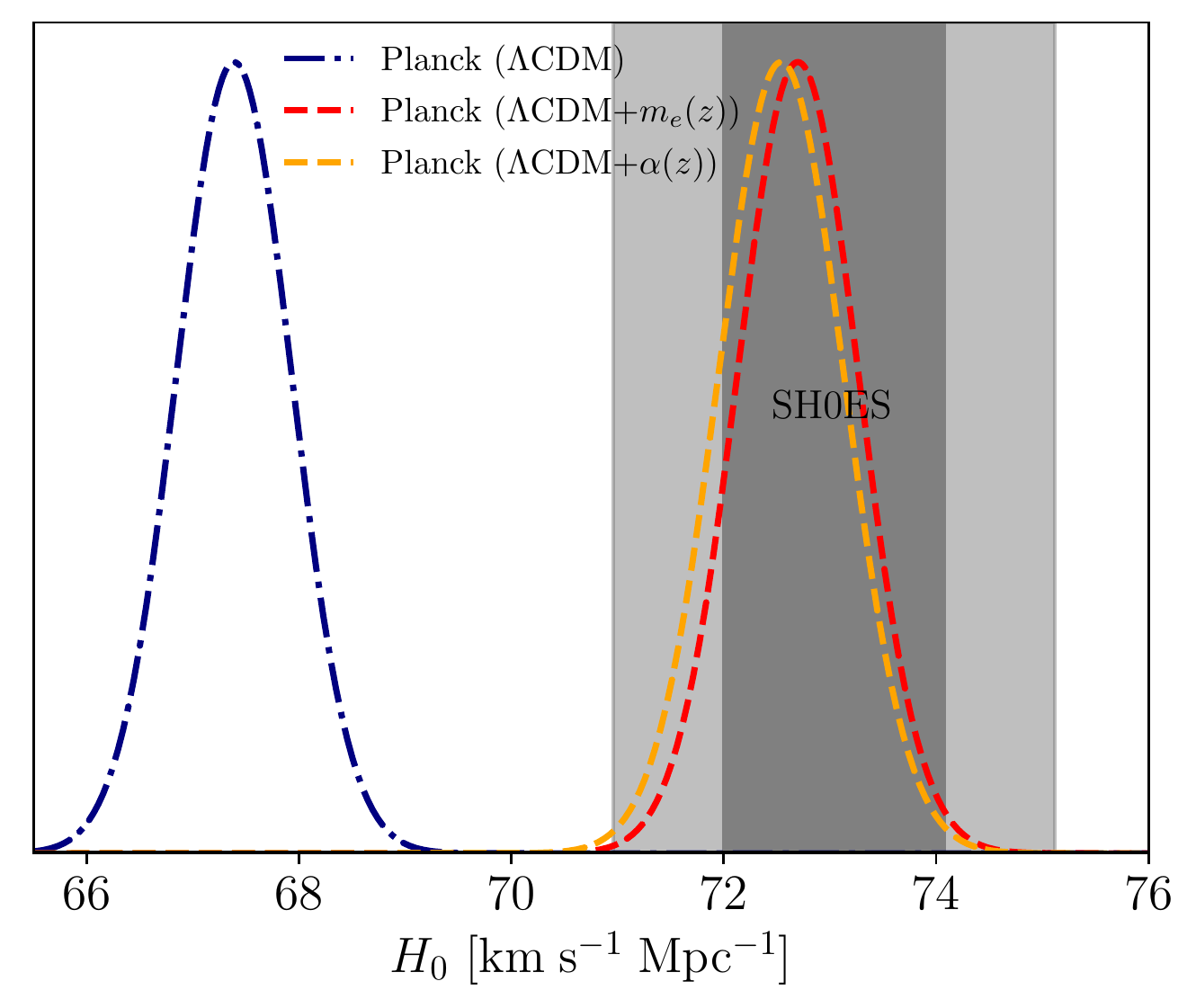}
\includegraphics[width = .322\columnwidth,trim= 00 17 00 30]{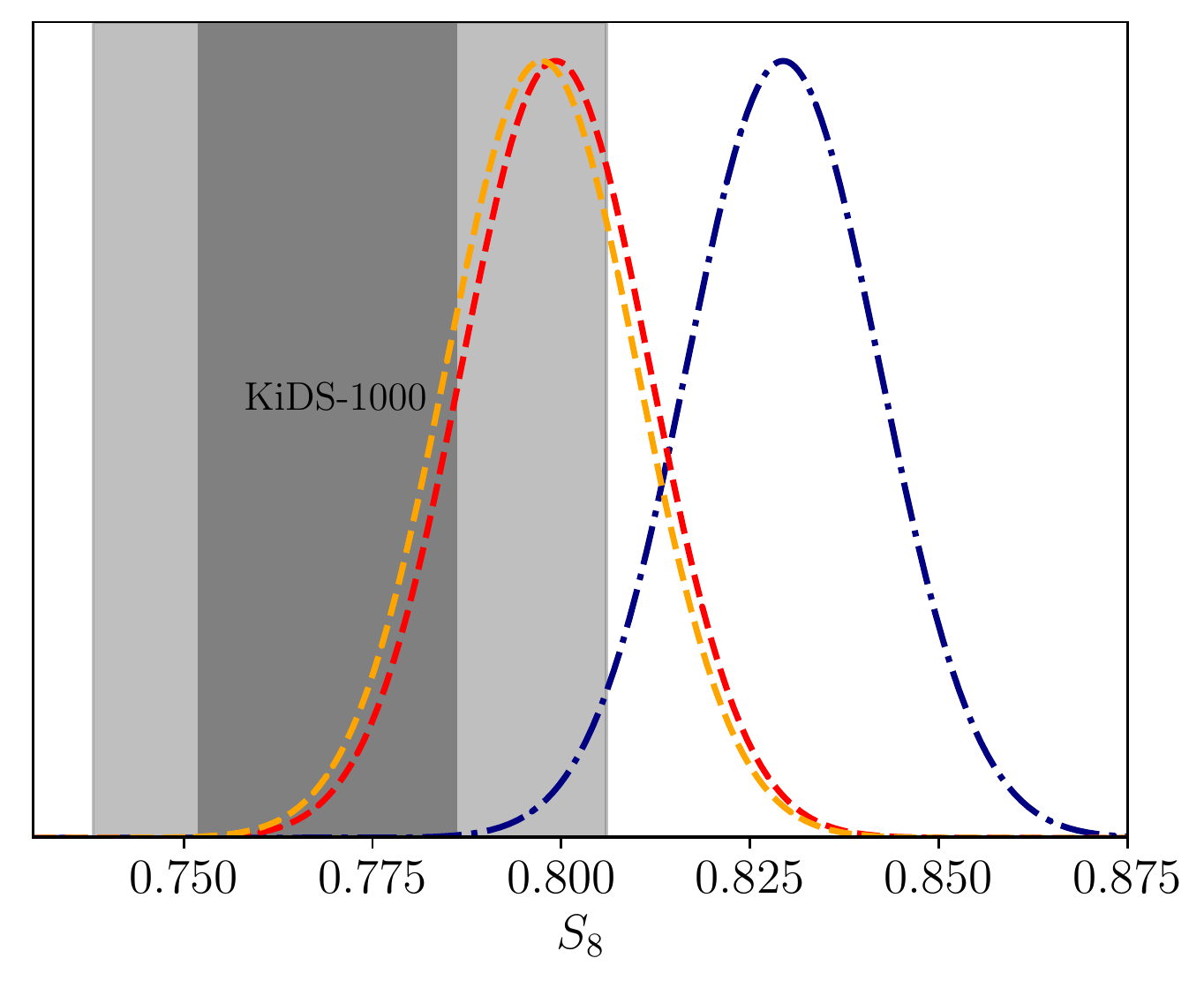}
\includegraphics[width = .325\columnwidth,trim= 00 17 00 30]{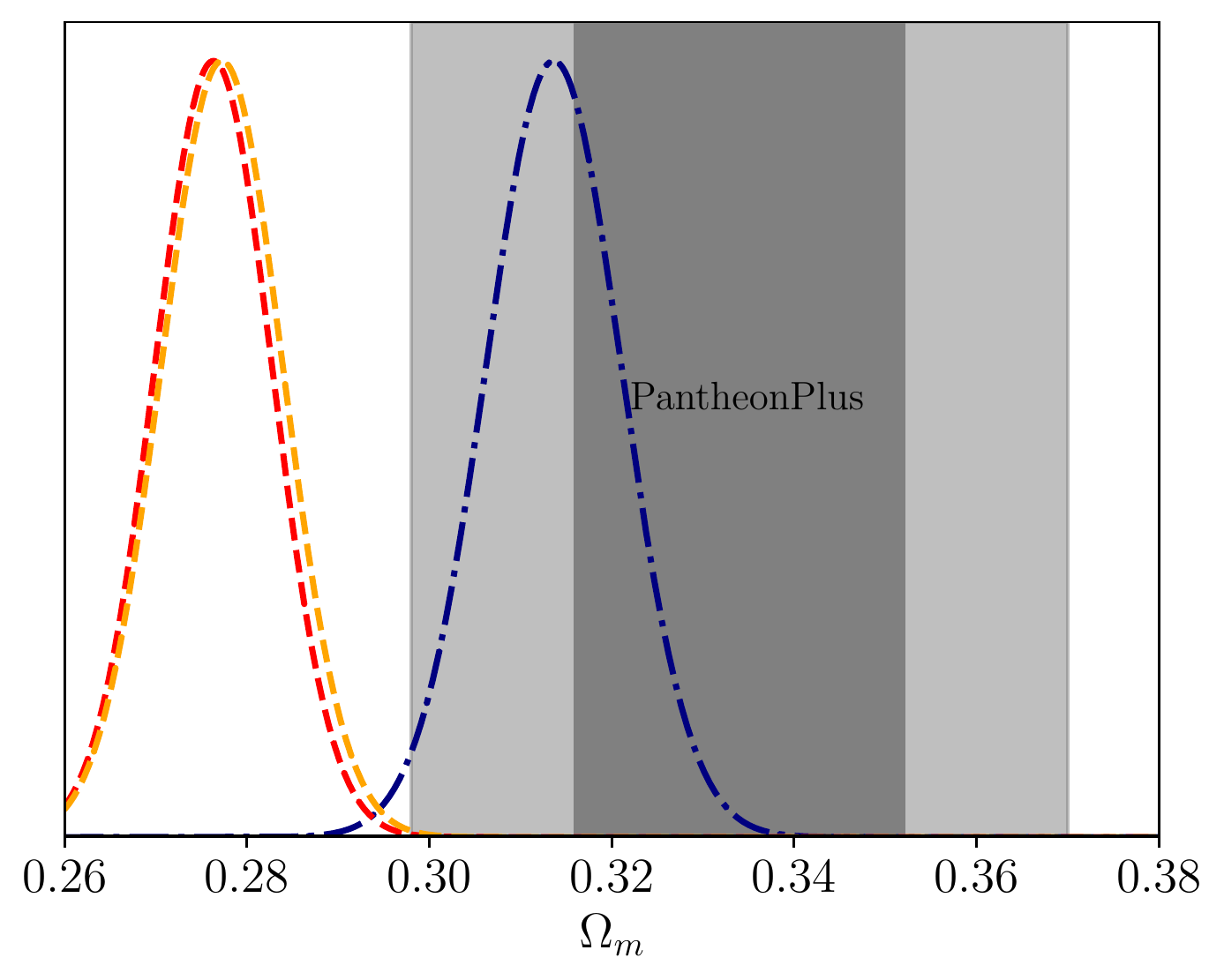}
\caption{The same plots as Fig.~\ref{fig:H0Om} but together with $\Lambda$CDM + $\alpha(z)$ model which is the time-varying fine structure constant obtained by our formalism with Planck CMB data (black curve in the left panel of Fig.~\ref{fig:alpha}).}
\label{fig:H0Om-alpha}
\end{figure}

\begin{figure}[H]
\includegraphics[width = .89\columnwidth,trim= 00 40 00 40]{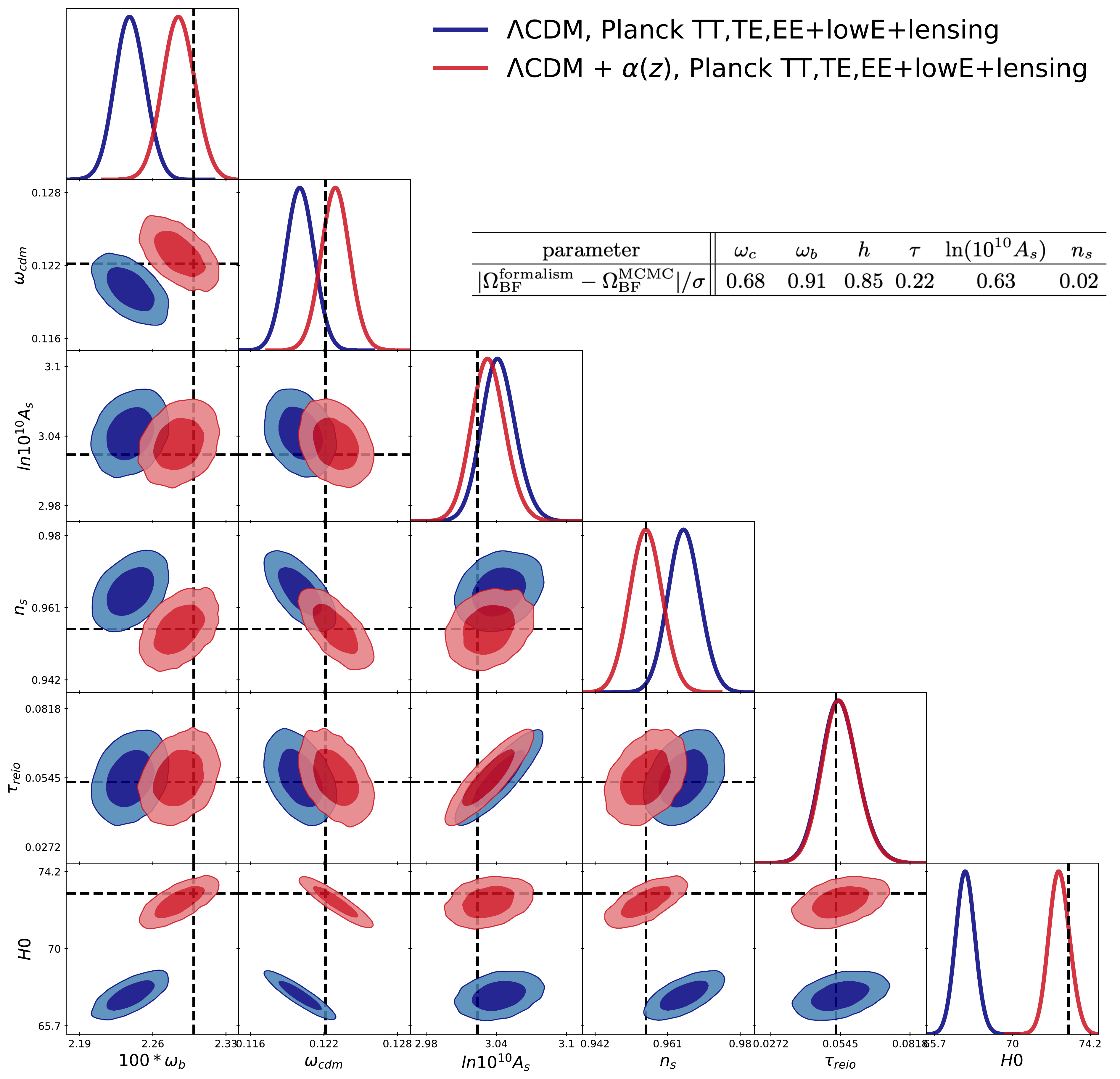}
\caption{Contour plot with two models, $\Lambda$CDM and $\Lambda$CDM + $\alpha(z)$. The latter model is what's found with Planck CMB data as a solution to the Hubble tension with SH0ES, which is shown as black curve in the left panel of Fig.~\ref{fig:alpha}. The black dashed lines are the estimated new best-fits by our formalism, Eq.~\eqref{eq:DObf_X}. Biases in terms of uncertainty of each parameter are shown in the inset table. The resulting constraints from $\Lambda$CDM + $\alpha(z)$ model are $H_0=72.54\pm0.59\;\text{km\;s}^{-1}\text{Mpc}^{-1}$ and $S_8=0.797\pm0.013$ with $\Delta \chi^2 = +0.62$. Note that the $\omega_b$ in the $\Lambda$CDM + $\alpha(z)$ model becomes less consistent with the BBN constraint from Ref.~\cite{Pisanti:2020efz} ($\omega_b = 0.0220 \pm 0.0005$) obtained using the helium abundance measurements from Refs.~\cite{Aver:2015iza,Peimbert:2016bdg,Hsyu:2020uqb}, but more consistent with those derived using the Helium mass fraction found by Ref.~\cite{Izotov:2014fga} ($\omega_b = 0.0234 \pm 0.0005$). Note that, however, while the constraints on the number of relativistic species $N_{\rm eff}$ \cite{Pisanti:2020efz} based on Ref.~\cite{Aver:2015iza,Peimbert:2016bdg,Hsyu:2020uqb} are consistent with the fixed value $N_{\rm eff}=3.046$ in our analysis, the constraint based on Ref.~\cite{Izotov:2014fga}, $N_{\rm eff}=3.60\pm0.17$, is not consistent with it.}
\vspace{-1\baselineskip}
\label{fig:MCMC-alpha}
\end{figure}

While we mainly present results with time-varying electron mass $m_e(z)$ in the main text, we consider a time-varying fine structure constant $\alpha(z)$ as well. In this appendix, we present all the equivalent results when we consider a time-varying fine structure constant $f(z)=\ln \alpha(z)$ instead of $\ln m_e(z)$. Figures~\ref{fig:deriv-Omega-alpha}--\ref{fig:MCMC-alpha} with $f(z)=\ln \alpha(z)$ are equivalent to those shown with $f(z)=\ln m_e(z)$ (Figs.~\ref{fig:me-cmb},\ref{fig:H0Om},\ref{fig:deriv-Omega},\ref{fig:deriv-chi2},\ref{fig:MCMC-me}, and \ref{fig:me-bao-pantheon}). Note that the conclusion is the same as that with $m_e(z)$ with almost the same shifts in best-fit parameters: a solution to the Hubble tension between Planck CMB \cite{Planck2018} and SH0ES \cite{Riess:2021jrx} can be found with a cost of worse fits to BAO \cite{BOSS:2016wmc} and PantheonPlus \cite{Brout:2022vxf}.

\newpage

The differences in the results of two extensions, time-varying electron mass $m_e(z)$ and fine structure constant $\alpha(z)$, can be understood according to the different dependencies of the energy levels of hydrogen and helium, atomic transition rates, photo-ionization/recombination rates summarized in the Appendix.~\ref{Appendix:numerical-techniques}. For example, the overall larger amplitudes of functional derivatives in Fig.~\ref{fig:deriv-Omega-alpha} and \ref{fig:deriv-chi2-alpha} compared to the case with $m_e(z)$ can be understood by the stronger dependence of $\alpha$ in Eq.~\eqref{eq:A2sA2p}-\eqref{eq:Teff}. Interestingly, however, the shapes are almost identical implying that the most important quantity, which determines the effect of non-standard electron mass and fine structure constant during hydrogen recombination is the effective temperature with which the hydrogen energy levels are calculated.

The other interesting difference is apparent in the obtained solutions at high redshifts ($z\gtrsim 1500$) in Fig.~\ref{fig:alpha} compared to Fig.~\ref{fig:me-cmb} and \ref{fig:me-bao-pantheon}. The high-redshift behaviors of solutions is related to Silk damping of which scattering rate is proportional to $n_e \sigma_T$. This explains the opposite behaviors of $m_e(z)$ and $\alpha(z)$ at high redshifts in Fig.~\ref{fig:me-cmb}, \ref{fig:me-bao-pantheon} and \ref{fig:alpha}, which are due to the opposite dependence of Thompson scattering cross section $\sigma_T$ on $m_e$ and $\alpha$, Eq.~\eqref{eq:sigmaT}.


One interesting result is that the $m_e(z)$ provides a better fit, which is mainly due to the better fit to Planck high-$\ell$ CMB spectra ($\Delta \chi^2=-1.15$ for $\Lambda$CDM + $m_e(z)$ and $\Delta \chi^2=+0.62$ for $\Lambda$CDM + $\alpha(z)$ model, respectively). One of the main challenges to fit high-$\ell$ spectra with an extension to $\Lambda$CDM is to preserve the Silk damping scale. For example, free-streaming dark radiation cannot keep a good fit to high-$\ell$ CMB spectra due to the increase of Silk damping and perturbation drag effect \cite{Bashinsky:2003tk,Lesgourgues:2013sjj,Follin:2015hya,Baumann:2015rya}. Unless the damping scale is preserved, any extension of $\Lambda$CDM model cannot provide a good fit to high-$\ell$ data. However, Ref.~\cite{Sekiguchi:2020teg} showed that the Silk damping scale can be preserved by adjusting the baryon energy density $\omega_b$ when there is a constant change in electron mass at recombination. Even though the situation is more complicated in our case since we have the value of the electron mass varying during recombination, this provides a partial explanation as to why time-varying electron mass provides a better fit to CMB data than time-varying fine structure constant and model-independent modifications to free-electron fraction (see also Ref.~\cite{Ge:2022qws} for a discussion of the CMB constraints on light relics due to this high-$\ell$ Silk damping tail).

\end{appendix}

\end{document}